# Operando Photonic Band Gap Probe of Battery Electrode Materials

Alex Lonergan[1], Umair Gulzar[1], Yan Zhang[1] and Colm O'Dwyer[1,2,3,4]*

[1]*School of Chemistry, University College Cork, Cork, T12 YN60, Ireland*

[2]*Micro-Nano Systems Centre, Tyndall National Institute, Lee Maltings, Cork, T12 R5CP, Ireland*

[3]*AMBER@CRANN, Trinity College Dublin, Dublin 2, Ireland*

[4]*Environmental Research Institute, University College Cork, Lee Road, Cork T23 XE10, Ireland*

## Abstract

**Innovative new materials are consistently emerging as electrode candidates from lithium-ion battery research, promising high energy densities and high-rate capabilities. Understanding potential structural changes, morphology evolution, degradation mechanisms and side reactions during lithiation is important for designing, optimising and assessing aspiring electrode materials. In-situ and operando analysis techniques provide a means to investigate these material properties under realistic operating conditions. Here, we demonstrate an operando spectroscopic method using photonic crystal-structured electrodes that uses the optical transmission spectrum to monitor changes to the state of charge or discharge during lithiation and the change to electrode structure, in real-time. Photonic crystals possess a signature optical response, with a photonic bandgap (or stopband) presenting as a structural colour reflection from the material. We leverage the presence of this photonic stopband, alongside its intricate relationship to the electrode structure and material phase, to correlate electrode lithiation with changes to the optical spectrum during operation. In this work, we explore the optical and electrochemical behaviour of a $TiO_2$ anode in a lithium-ion battery, structured as an inverse opal photonic crystal. In principle, the operando technique demonstrated here is versatile and applicable to a wide range of electrochemical electrode material candidates when structured with ordered porosity akin to a photonic crystal structure.**

# Introduction

The growing popularity of smartphones, electric vehicles and wearable electronic devices, alongside their associated high energy demands, has driven the research for lithium-ion battery (LIB) materials towards achieving higher energy densities, higher cycling stability, longer lifespans, safer battery cells, lighter and more flexible designs[1,2,3]. Future energy storage devices require highly energy dense materials which can be manufactured into versatile designs to support emerging technological innovation[4,5]. Research attempts to facilitate these requirements commonly explore alternative electrode materials for both the anode[6,7] and the cathode[8,9]. Another approach focuses on exploring novel geometries or material dimensions, with particular emphasis on nano-sized architectures, for the electrode materials[10,11]. Incorporating nano-sized features into the electrode design has reported advantages of shorter ion diffusion lengths with a larger surface area exposed, improved rate capability from the increased electrolyte/electrode interface and better accommodation of material volume changes due to inherent porosity or particle spacing[12,13].

Mainstream commercial uptake of these alternative electrode materials or designs requires detailed information on side reactions, degradation mechanisms, particle cracking, morphology evolution, material phase changes and state of charge heterogeneity in order to assess the health and long term stability of these potential electrodes in functioning cells[14,15,16]. In-situ and operando characterisation techniques offer pathways to assess the battery performance and electrode material evolution non-invasively, allowing for this information to be obtained under realistic operating conditions by maintaining the sealed battery environment during analysis[17,18,19] This is contrary to ex-situ or post-mortem analytical techniques which deconstruct the battery cell for analysis, introducing the possibility of material contamination which may not provide an accurate description of performance[20].

Development and progress related to in-situ and operando techniques in the literature has taken many forms, cleverly utilising a variety of different analytical techniques to probe aspects of LIB behaviour. In-situ x-ray diffraction (XRD) has been used to track structural changes and reversibility in anatase $TiO_2$ electrodes[21] and assess the crystallinity and identify metastable phases in silicon nanowires[22]. In-situ and operando Raman spectroscopy can monitor the intensity and position of phonon modes which has been used to identify structural phase transitions (or lack thereof) in TiO$_2$ nanowire and nanoparticle electrodes[23] or pinpoint the lithiation potential in doped silicon anodes using the transition from crystalline to amorphous silicon as an indicator[24]. Operando atomic force microscopy (AFM) has been used to study dendrite growth on lithium metal electrodes[25] or assess the mechanical properties of silicon anodes during lithiation processes[26]. A myriad of other techniques used in the literature for in-situ and operando studies include nuclear magnetic resonance (NMR)[27], scanning electron microscopy (SEM)[28], transmission electron microscopy (TEM)[29], fourier transform infrared (FTIR) spectroscopy[30] and acoustic ultrasound transmission[31], to name just a few. Critically, each of these techniques offers an insight into LIB operation which cannot be obtained through conventional ex-situ methods. Recently, operando and in-situ optical characterisation techniques have begun to emerge. Optical scattering intensity from electrode materials and particles has been used to track phase transitions and state of charge in specific materials[32] [33]. Optical fibre Bragg grating sensors have been embedded in coin cells in efforts to provide non-invasive operando assessment of chemo-mechanical stress in electrode materials[34]. Cross-sectional optical microscopy images of lithium-ion full cells have been used to provide in-situ insights into lithiation-linked color changes and electrode thickness evolution[35]. In-situ UV-Vis spectroscopy has been shown to provide useful supplementary characterisation to electrochemical processes by determining charge storage mechanisms and tracking of changes in material oxidation states[36]. Adopting optical probes provide many opportunities to evaluate mechanical/volumetric changes, interfacial species changes, electrolyte composition and absorption characteristics related to electronic processes with materials, to name a few possibilities. While invasive probes prove useful to evaluation cell-level state of health in batteries, there are challenges in assessing how active material interconnection and changes can be tracked in real-time, a problem that lies between spectroscopic identification of surface species or changes and electrode-level response.

Here, we take an approach to develop an operando electrode characterisation technique that exploits the unique characteristics of photonic crystals (PhCs), where the periodicity in material refractive index creates a pseudo photonic band gap, somewhat analogous to the electronic band gap in semiconducting materials, to disallow propagation of certain frequencies of light. More specifically, we leverage the presence of the photonic stopband in inverse opal (IO) structured TiO$_2$ anodes to monitor the behaviour of the electrode non-destructively during reversible lithitation-induced changes to the interconnected active material. Electrodes incorporating an IO PhC design have been explored to a large degree in the literature, and serve as model electrodes for investigating porosity and the influence of interconnection between active materials without binders or additives[37][38][39]. However, these IO PhCs are much more than just a simple structural template and are renowned for their ability to manipulate specific frequencies of light[40][41]; the periodic dielectric structure establishes a 'structural colour' reflection from the PhC surface[42][43]. The structural colour inherent to these materials is a key component to the unique potential of the technique proposed here. Electrodes which adopt a periodic PhC can exploit this observable structural colour to provide information on battery material behavior, non-destructively and without specialised equipment, provided the battery cell was constructed to facilitate observation of the electrode material. Recently, the shrinking and swelling of PhC hydrogels have used changes in structural colour to track water content present in solid-state electrolyte in zinc-air batteries[44], demonstrating the capability of a PhC-based optical monitoring system.

While ordered and interconnected porous materials such as inverse opals have show very good stable response in LIBs using a range of cathode and anode materials, there remain some opens questions on how such electrodes vary during cycling since it has been assumed that porosity (acknowledging the gravimetric and volumetric energy density penalty) is believe to facilitate the typical volumetric change, while the interconnected nature maintains electrical conductivity[41]. But deformation accommodation mechanically, must also occur if that is the case, or a change to periodicity must accompany large volume changes. Similar queries arise for random porous materials and the nature of material packing, ionic and electronic conductivity and the benefits of non-tortuous porosity. The marriage of photonic crystal and electrochemical and physical processes is examined here as a operando tracking probe to determine how they behave during cycling. The idea for this concept has been proposed for some time[45], though there has yet to be a practical implementation for the technique. In this work, we showcase a new operando characterisation technique for LIBs which directly exploits this phenomenon. Our results suggest that for $TiO_2$ IO electrodes, the wavelength position and transmission intensity of the photonic stopband is intimately linked to the degree of lithiation of the electrode. The degree of light transmission intensity, as evidenced by operando spectra, is found track with the operating voltage for both charge and discharge processes; a cyclical variation in the optical transmission emerges during charging and discharging. Operando transmission spectra recorded during cyclic voltammetry and galvanostatic charge/discharge cycles of $TiO_2$ IO electrodes exhibit consistent red-shifting of the photonic stopband energy during cycling. We discuss the significance of these results, correlating the shifts in the photonic stopband to the physical properties of the IO electrode. Most notably, the operando technique explored here is not restricted to just $TiO_2$ IO electrodes; the method is versatile and applicable to a wide range of electrochemically active electrode materials that can be structured to feature a photonic stopband.

## Defining the conditions for operando analysis of $TiO_2$ IO anodes

The optical probe is based on an angle-resolved transmission measurement arrangement (Figure 1(a)) where the $TiO_2$ IO electrode is measured while charging and discharging in a three-electrode cell (Figure 1(b)). Here, we analyse $TiO_2$ IO anodes coated with a thin layer of Ni metal to provide electrical conductivity of a current collector while maintaining optical transparency. More detailed information on the choice of materials and electrode design can be found in the Methods section. When dealing with the optical response of PhCs, an important consideration for the location of the photonic stopband is the periodicity of the structure. For IOs, the centre-to-centre pore distance is often used as an indicator of structural periodicity. For our $TiO_2$ IOs, we measured centre-to-centre pore distances (Figure 1(c)) and they vary in the range 320 – 330 nm. These $TiO_2$ IOs were prepared via sol-gel infiltration of PS spheres (see Methods section for further details) with initial diameters in the range 370 – 380 nm, exhibiting a well-documented uniform shrinkage upon inversion to the IO[42,43,46]. A typical SEM image of a $TiO_2$ IO following the deposition of a 20 nm Ni metal layer can be seen in Figure 1 (e); the deposition of the metal layer through magnetron sputter coating has the effect of thickening the top surface of the IO walls. Additional SEM images for the $TiO_2$ IO post-cycling can be found in the Supplementary Materials Fig. S2. These $TiO_2$ electrodes are materials we have investigated for many years and in binder-free and conductive-additive free formulations, they provide very stable multi-thousand cycle life response in standard carbonate electrolytes and offer a reliable and well characterised LIB anode system to carefully extract a correlation between spectroscopic and electrochemical response. There are some open questions such as how anatase $TiO_2$ responds to lithiation during a cycle and over long cycling lifetimes, how interconnected porous materials generally behave during initial and longer term cycling in terms of material expansion and the supposed intrinsic benefits of porosity and what changes occur in materials that are relatively stable of long cycle durations.

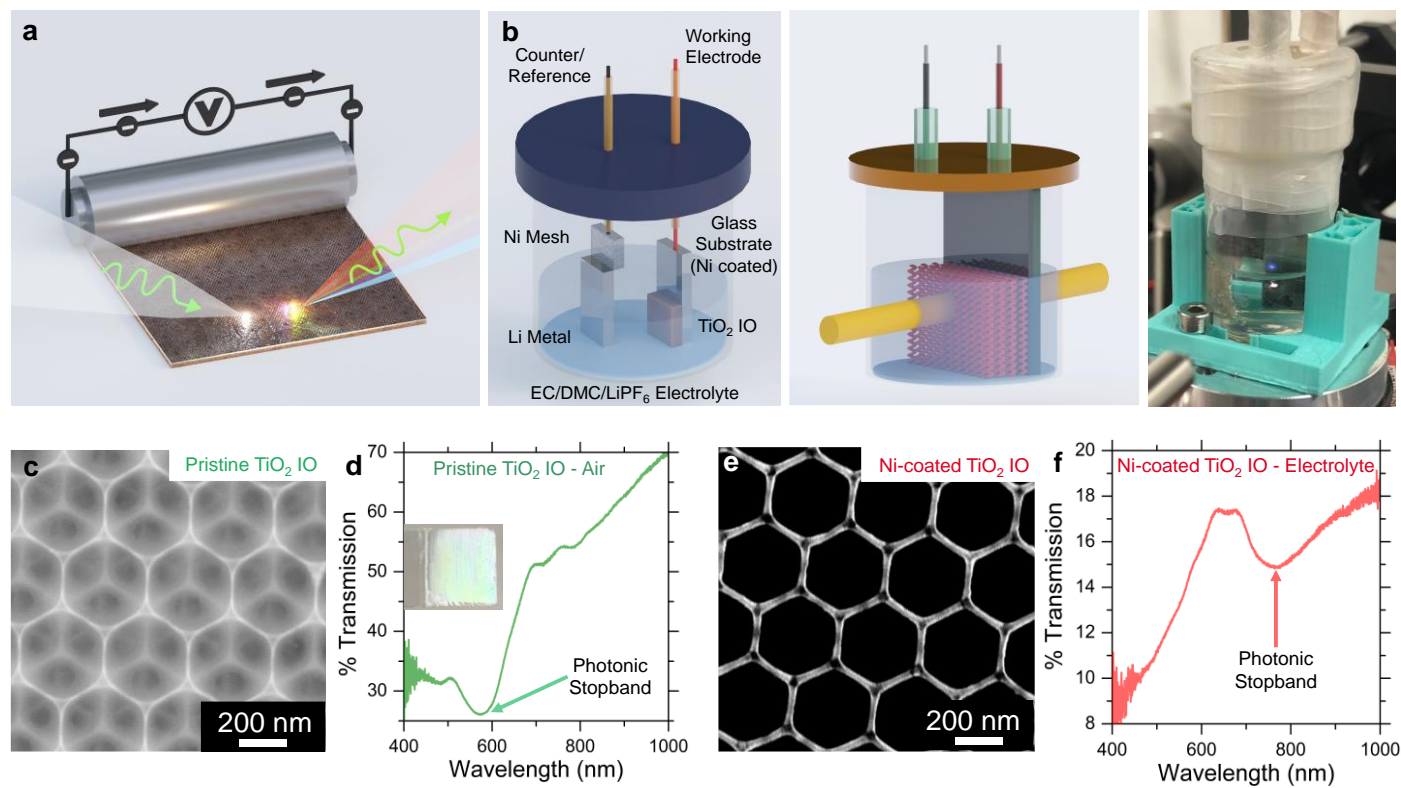

**Figure 1 Operando spectroscopy and photonic crystal battery electrodes.** (a) The operando technique for photonic crystal LIB electrodes obtains simultaneous optical spectra and electrochemical charge-discharge data from the ordered macroporous $TiO_2$ inverse opal electrode. (b) The battery cell uses a three-electrode configuration to allow light transmission through the electrode. Lithium metal is used as the counter electrode and reference electrode, with sputtered Ni forming the current collector to an underlying glass substrate to the working electrode. (c) Scanning electron microscopy of a typical $TiO_2$ IO used as an electrode prior to any modifications shows the highly ordered interconnected nature of the material. (d) The optical transmission spectrum of a typical unmodified $TiO_2$ IO prepared using 350 nm polystyrene opal spheres as a sacrificial template, showing a photonic stopband centred at approximately 573 nm. (e) SEM image of a typical $TiO_2$ IO following magnetron sputter deposition of 20 nm of Ni metal onto the surface to form the current collector. (f) A $TiO_2$ IO following deposition of 20 nm of Ni metal and immersion of the IO into ethylene carbonate:dimethyl carbonate:$LiPF_6$ battery electrolyte, showing a red-shifted photonic stopband centred at ~766 nm.

Optical transmission spectra for typical $TiO_2$ IOs used in this study are shown in Figs. 1 (d) and (f) for a pristine $TiO_2$ IO in air and a Ni-coated $TiO_2$ IO in $LiPF_6$ electrolyte, respectively. The photonic stopband for the IO in air at 572 nm red-shifts to 766 nm upon infiltration with electrolyte. This effect is expected due to the increase in the effective refractive index of the IO when the electrolyte replaces the air in the porous structure[42 46]. The Bragg-Snell model is commonly used to estimate and predict the central position of photonic bandgaps and stopbands. At 0° incidence, the minimum transmission wavelength $\lambda_{111}$ for light reflected from the primary (111) plane is found from the Bragg-Snell relation using the effective refractive index $n_{eff}$ and the interplanar spacing $d_{111}$ as:

$$\lambda_{min} = 2\, d_{111}\, n_{eff} \tag{1}$$

A common method, of which there are a few, used to calculate the $n_{eff}$ is based on a volume averaging of the IO refractive index $n_{IO}$, with volume fraction $\varphi_{IO}$ (typically, $\varphi_{IO} = 0.26$), and the background filling material refractive index $n_{bg}$, with volume fraction $\varphi_{bg}$ (typically, $\varphi_{bg} = 0.74$), as:

$$n_{eff} = n_{IO}\, \varphi_{IO} + n_{bg}\, \varphi_{bg} \tag{2}$$

Our previous work[43 46 47] with anatase TiO$_2$ IOs found a good agreement between effective refractive index estimates using eqn. (2) and experimentally measured optical transmission spectra for $n_{IO} = 2.488$, yielding an estimate for the effective refractive index of $n_{eff} = 1.387$. For the primary (111) reflection plane, $d_{111}$ can be calculated for ideal and isotropic PhCs using the centre-to-centre pore distance $D$ via the relation:

$$d_{111} = \sqrt{\frac{2}{3}}\, D \tag{3}$$

Measurements with TiO$_2$ IOs show that a reduced interplanar spacing is present in TiO$_2$ IOs prepared this way[43 46], possibly attributed to anisotropic shrinking of the space between layers compared to the pore sizes on the horizontal plane. Importantly for this work, any changes to the optical spectra from changes to material dielectric constant, electrolyte or thickness/periodicity are comparative to an initial state or position, allowing for stopband shifts to be simply related to changes in $n_{eff}$ or $d_{111}$, as in eqn. (1).

## Reversible lithiation that causes cyclic fluctuations in optical transmission

Since the approach is sensitive to changes in both material refractive index and periodicity or dimension, the fundamental electrochemical response was defined at the outset. Figures 2 (a) and (b) show the voltammetric response observed on a typical cyclic voltammogam (CV) for a TiO$_2$ IO electrode for the first and second cycle, respectively. The current peaks at ~ 1.63 V in the cathodic sweep and ~ 2.20 V in the anodic sweep are typical of reversible lithiation reactions with anatase phase TiO$_2$[48], peaks present just below ~ 1.3 V in the cathodic sweep may suggest the presence of a bronze phase $TiO_2$(B), potentially indicating a mixture of both phases[49]. However, there is a large peak at ~ 2.20 V in the cathodic sweep of the first CV cycle which is not present in the second cycle, or any subsequent cycles for that matter. The peak voltage for this reaction coincides with the reported reaction potential for the conversion of Ni metal to NiO[50]; the reverse reaction (NiO conversion to Ni metal) is not present in this potential window (1.0 – 3.0 V) as this reaction is reported to happen below 1.0 V[50] and so the modification of the Ni current collecting deposit remains unique to the very first cycle.

The operando monitoring of the $TiO_2$ IO electrode immediately revealed some significant effects on the appearance of the transmission spectra, compared to the initial spectra prior to cycling. One such effect is visible in Figures 2 (c) and (d), which shows several transmission spectra, recorded at 3.0 V, for two different $TiO_2$ IO electrodes undergoing cyclic voltammetry (Figure 2 (c)) and galvanostatic charge/discharge (Figure 2 (e)). For every electrode we cycled, and every electrochemical process we tested, we find a significant irreversible drop in transmission intensity through the electrode, which occurs after the first cycle and, critically, remains essentially unchanged for all subsequent cycles. The oxidation of the Ni metal current collector layer to NiO is most likely the underlying cause for the significant reduction in transmission intensity unique to the first cycle; the conversion process ceases after the first cycle, with no substantial drops in transmission intensity observed after this point. The photonic stopband also undergoes a considerable red-shift (~16 nm) in the first cycle between 3.0 V and 2.0 V, likely linked to the conversion of Ni to NiO and subsequent increase in effective refractive index of the PhC[47], see Supplementary Materials Fig. S3 for more details.

Intercalation-mode Li-ion battery material involve reversible lithium insertion and removal during each cycle. This often results in volumetric swelling, sometimes causing cracking in cyclic swelling/contraction cycles, among other effects. Determining the relative contributions of composition (thus refractive index) changes and physical (dimension) changes is difficult and they are often measured separately. Our approach couples both parameters and we find that real-time monitoring of these processes exhibits cyclic behaviour coincident with the charging and discharging cycles. Figures 2 (e) and (f) show a series of operando transmission spectra recorded at 0.1 V intervals for a typical cathodic sweep (discharge, 3.0 V – 1.0 V) and the corresponding anodic sweep (charge, 1.0 V – 3.0 V), respectively. Most notably, we see a cyclic fluctuation in the optical transmission intensity that is closely linked to the operating voltage. Lithium insertion into the electrode during the discharge acts to decrease the overall transmission intensity. The transmission intensity is then recovered to the initial intensity, at the cycle's beginning, when lithium is removed during the charge process.

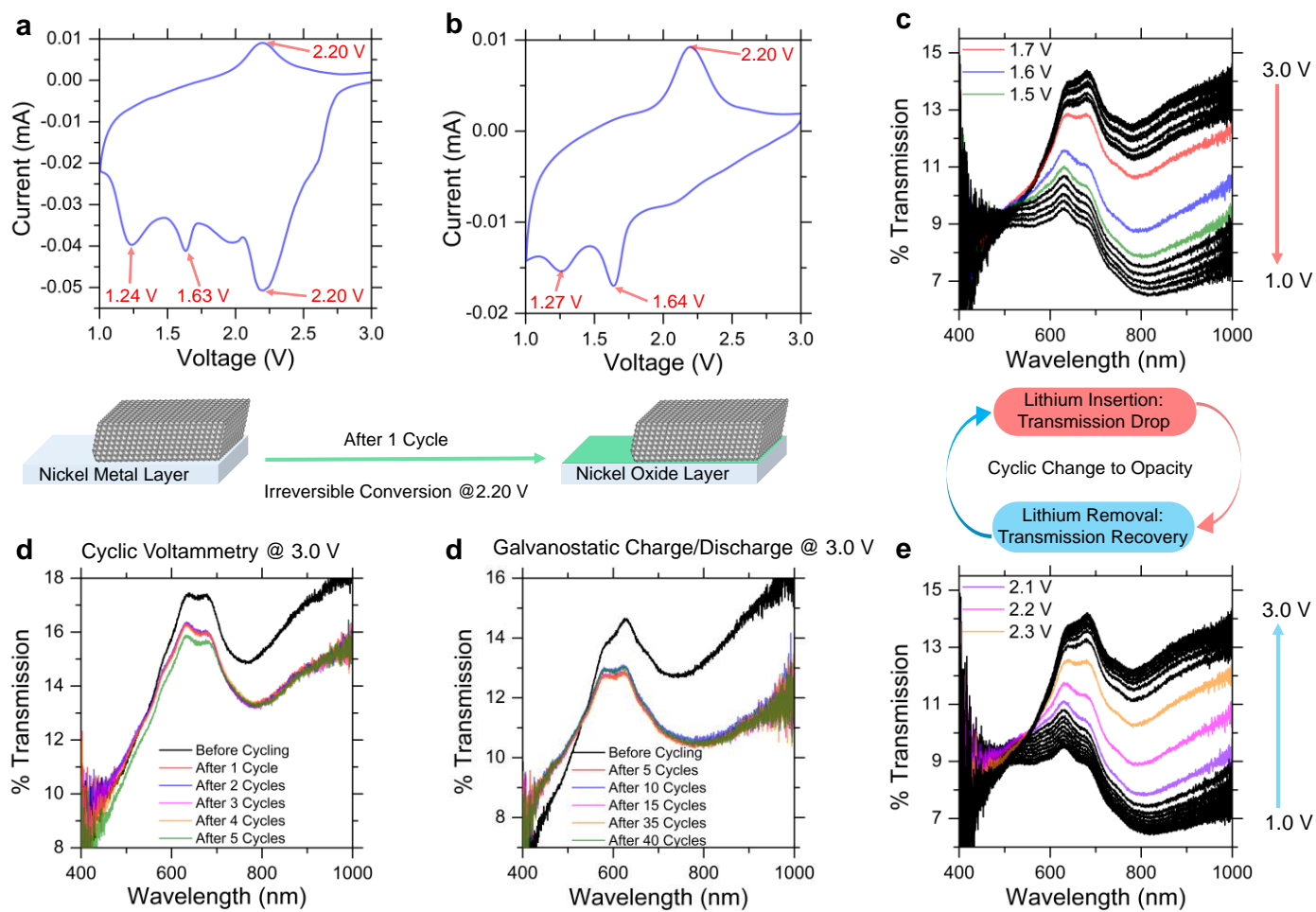

**Figure 2 Cyclic optical opacity and lithiation.** Labelled peaks on cyclic voltammetry curves for the (a) 1st and (b) 2nd cycles of a typical $TiO_2$ IO anode, cycled at 0.2 mV/s, showing the disappearance of a prominent peak centred at 2.20 V after the first discharge with a suggested link to an irreversible conversion of Ni to NiO in the 3 – 1 V voltage window. Operando optical spectra recorded for $TiO_2$ IO anodes at 3.0 V for electrodes undergoing (d) cyclic voltammetry and (e) galvanostatic charge/discharge processes, highlighting an initial significant irreversible drop in transmission intensity occurring in the early stages of battery life. Reversible and cyclic transmission intensity fluctuations shown via operando optical spectra recorded at 0.1 V intervals in a typical (c) discharge (cathodic sweep) and (d) charge (anodic sweep) for a $TiO_2$ IO anode. Certain spectra are accentuated in colour to highlight specific voltage intervals with the largest fluctuations in transmission intensity.

Here, this effect is shown for a typical CV cycle (0.2 mV/s) but we observe this cyclic process in every cycle during voltammetric measurements and also during galvanostatic charge-discharge cycles. The transmission intensity constantly, and reversibly, responds to the operating voltage and thus, the level of lithiation of the electrode i.e. the mole fraction of Li in Li$_\chi$TiO$_2$. This identifies an effective refractive index change and not a change to the dimensions of the TiO$_2$ electrode during lithium uptake or removal. Several of the operando spectra in Figures 2 (e) and (f) are displayed in colour to draw attention to certain spectra at specific voltages, namely, the spectra which coincide with the largest suppression or recovery in transmission intensity. In the discharge in Figure 2 (e), the largest drop in transmission intensity is found between 1.7 – 1.5 V. Likewise, during charging the most pronounced recovery in the transmission intensity is found in the operando spectra recorded between 2.1 – 2.3 V. Considering that for TiO$_2$, the lithium insertion reaction in the discharge occurs at ~ 1.63 V and the lithium removal reaction occurs at ~ 2.20 V, as per the CV data, there appears to be a strong correlation between the expected mole fraction of lithium content in the electrode and the transmission intensity. Greater lithium mole fractions in the electrode clearly cause a suppression of total optical transmission. The drops in transmission intensity are consistent across multiple cycles. Cyclic opacity in the electrode is thus linked to the state of charge or depth of discharge. For example, during the lithiation of the electrode in the discharge process, voltages between 1.7 – 1.5 V are the onset of more optically opaque electrodes and this voltage window in a typical galvanostatic discharge process coincides with an increase in the lithium mole fraction in Li$_\chi$TiO$_2$ from $\chi$ ~0.1 to $\chi$ ~0.25.

**The photonic band-gap and changes to material dimensions in operando**

Beyond the overall optical transmission intensity changes during cycling, which track the reversible lithiated state of the electrode, spectral features near the photonic band-gap of the ordered macroporous electrode can reveal much finer changes in the optical profile and so the dimensions and changes to the TiO$_2$ IO. As per eqns. (1) – (3), the photonic stopband of PhC materials is inherently sensitive to the IO pore spacing and refractive indices of the materials comprising the porous structure. Figure 3 (a) shows the calculated specific capacity at cycle 7 and cycle 70 for a TiO$_2$ IO undergoing repeated galvanostatic charge/discharge processes at a C-rate of 2C, showing a respectable performance with a discharge specific capacity of ~ 160 mA h g$^{-1}$ at that fast rate, which comparing favourably with other reports for $TiO_2$ materials[38][39] and our previous examination of these types of electrodes over 5000 cycles; the volumetric expansion is small in the first several hundred cycles. Calculated mole fractions of lithium incorporation into $Li_\chi TiO_2$ based on the theoretical specific capacity for $TiO_2$, that is 168 mA h g$^{-1}$ for $Li_{0.5}TiO_2$ or 335 mA h g$^{-1}$ for $Li_1TiO_2$[23][39], indicate that ~0.5 moles ($Li_{0.5}TiO_2$) are incorporated at full discharge, a common occurrence with $TiO_2$ electrodes. Figure 3 (b) shows the corresponding optical transmission spectra for cycle 7 and cycle 70, both recorded at 3.0 V. An enlarged view of the photonic stopband is shown in Figure 3 (c), showing the subtle changes in the stopband energies during cycling. Most importantly, the shift in the central wavelength position of the photonic stopband is emphasised with ~10 nm red-shift observed between cycle 7 and 70. Additional operando optical spectra for these measurements, showing the cyclic reversible transmission intensity, can be found in the Supplementary Materials Fig. S4.

Figure 3 (d) depicts the measured stopband minimum, recorded from operando spectra at 3.0 V, versus the number of completed full (charge an discharge) cycles. Interestingly, there appears to be a linear relationship between the stopband position and the number of completed cycles; the cycle life of the electrode is indicated by the red-shift in stopband position. Over extended (several hundred to several thousand) cycling periods, the structural walls of $TiO_2$ IOs have been observed to thicken[39] while retaining the periodicity and interconnected order. Quantifying this change in the IO structure as relating to an expansion of the interplanar spacing of the PhC or an increase in the effective refractive index of the composite material should allow for a correlation to be established between the observed structural colour of the material and the cycle number. Expansion occurs in all directions and the change to the stopband energy is linked to the (111) planes or the ordered crystal, i.e. its thickness. The electrochemical behaviour of the $TiO_2$ IO is stable across a number of cycles, as seen in Figure 3 (e) with consistent specific capacity values (~ 150 – 160 mA h g$^{-1}$). The Supplementary Material Fig. S5 provides more detail on the coulombic efficiency for this cell. The red-shift of the photonic stopband is a gradual and consistent process that occurs throughout the battery life, as illustrated in Figure 3 (f). Changes to the photonic stopband position are minor, often just a few nanometres of spectral shift, depending on the number of cycles completed. It is important to remember that the cyclic and reversible change in light transmission from reversible lithiation *during* each cycle in Figure 2 of course happens here also, while the spectrum red-shifts over extended cycling, and the changes to the dimension of the TiO$_2$ become apparent.

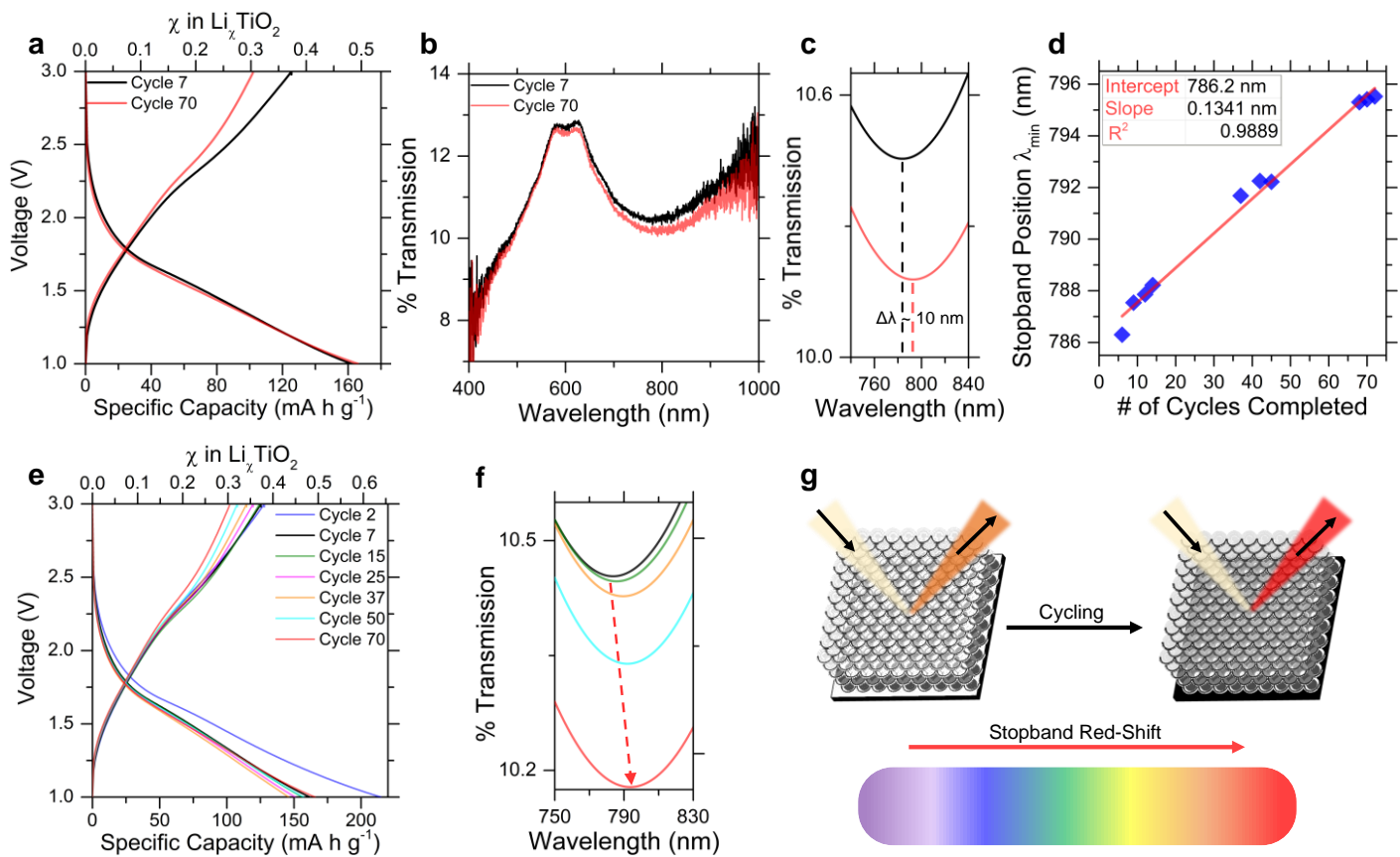

**Figure 3. Photonic stopband spectral shift during reversible lithiation.** (a) Voltage versus specific capacity for a typical TiO$_2$ IO anode undergoing galvanostatic charge/discharge cycling at 2C, showing cycle 7 and cycle 70. (b) The corresponding operando optical transmission spectra, both recorded at 3.0 V. (c) The same optical spectra with a closer focus on the gradual red-shift in stopband position with increased battery cycling, in this case highlighting a 10 nm red-shift recorded between cycle 7 and cycle 70. (d) A plot showing the recorded stopband position at 3.0 V versus the number of completed cycles for a TiO$_2$ IO anode at a charge/discharge rate of 2C. A linear fit is applied to the data, suggesting a predictable red-shift with increasing cycle number. (e) Recorded specific capacity values across a number of cycles in the galvanostatic charge/discharge process of a TiO$_2$ IO anode. (f) Enlarged view of the photonic stopband of a TiO$_2$ IO anode showing the gradual red-shift in the central stopband position at 3.0 V with higher cycle number. (g) A schematic diagram showing the proposed effect of battery cycling on the photonic stopband of IO structured electrodes, with a red-shift in the optical signature observed for increased cycling.

This gradual red-shift in stopband position is not limited to galvanostatic charge/discharge processes; it can also be observed in the repeated voltammetric cycling. Optically, the operando transmission spectra, recorded at 3.0 V in each case, for a number of selected cycles are shown in Figures 4 (a) – (c) for scan rates of 0.1, 0.2 and 0.3 mV/s, respectively. For each scan rate, the spectra are compared at 3.0 V after a period of 4 full cycles has passed. Importantly, there is a gradual red-shift in the stopband position with cycle number, observed in each case. At 0.1 mV/s the stopband shifts ~11 nm from 781 nm to 792 nm between cycle 1 and cycle 5. A smaller stopband shift of ~9 nm from 781 nm to 790 nm between cycle 1 and cycle 5 is observed for the electrode cycled at 0.2 mV/s. Finally, the smallest stopband shift of ~8 nm from 778.5 nm to 786.5 nm is observed for the electrode cycled at 0.3 mV/s between cycle 2 and 6. Figures 4 (d) – (f) show the recorded CV curves for the 2$^{nd}$ cycle for the three different TiO$_2$ IO electrodes at scan rates of 0.1, 0.2 and 0.3 mV/s, respectively. In each case, the most prominent reaction peaks are located at ~ 1.6 V in the cathodic sweep and ~ 2.2 V in the anodic sweep, typical of TiO$_2$ electrodes[48,49].

By examining the red-shifts, recorded after 4 completed cycles for each scan rate shown in Figures 4 (a) – (c), a trend in the optical data emerges. The largest red-shift of 11 nm is associated with the slowest scan speed (0.1 mV/s) and the smallest red-shift is linked to the quickest scan rate (0.3 mV/s). Extending this analysis to include the stopband red-shift data for additional CV cycles, shows the trend is maintained, as depicted in Figure 4 (g). The highest rate of change in the stopband position with cycle number is observed for the slowest scan speed (0.1 mV/s), featuring the steepest slope of 2.86 nm/cycle, with the converse statement being true for the slowest scan speed with a slope of just 1.60 nm/cycle. Since the potentiodynamic voltammetry does not force a constant reaction rate from a fixed current of a galvanostatic cycle, the fastest scan rate thus corresponds to the shortest lithiation duration. Additionally, the change in scan rate is just 0.1 mV/s, yet we observe a specific difference in response correlated to the degree of lithiation, showing how sensitive the measurement and lithiation processes are to voltage changes. The extended lithiation period of the overall cycle associated with slower scan speeds may contribute to this effect, with higher degrees of lithiation and repeated cycling causing further red-shifting of the stopband. By comparison to the galvanostatic data in Figure 2 (d), we note that the fixed current (reaction rate) under standard charge-discharge conditions results in a stopband shift of ~0.134 nm/cycle, slightly over an order of magnitude less that under voltammetric polarization in spite of a near identical effect of cyclic optical transmission variation during each cycle. Since our method is not susceptible to differences between the cumulative processes that contribute to the total integrated time in galvanostatic vs voltammetric cycling, the difference in stopband shift from changes to material dimension indicates that the galvanostatic data results in a lesser degree of lithiation-induced swelling per cycle than voltammetry under similar conditions.

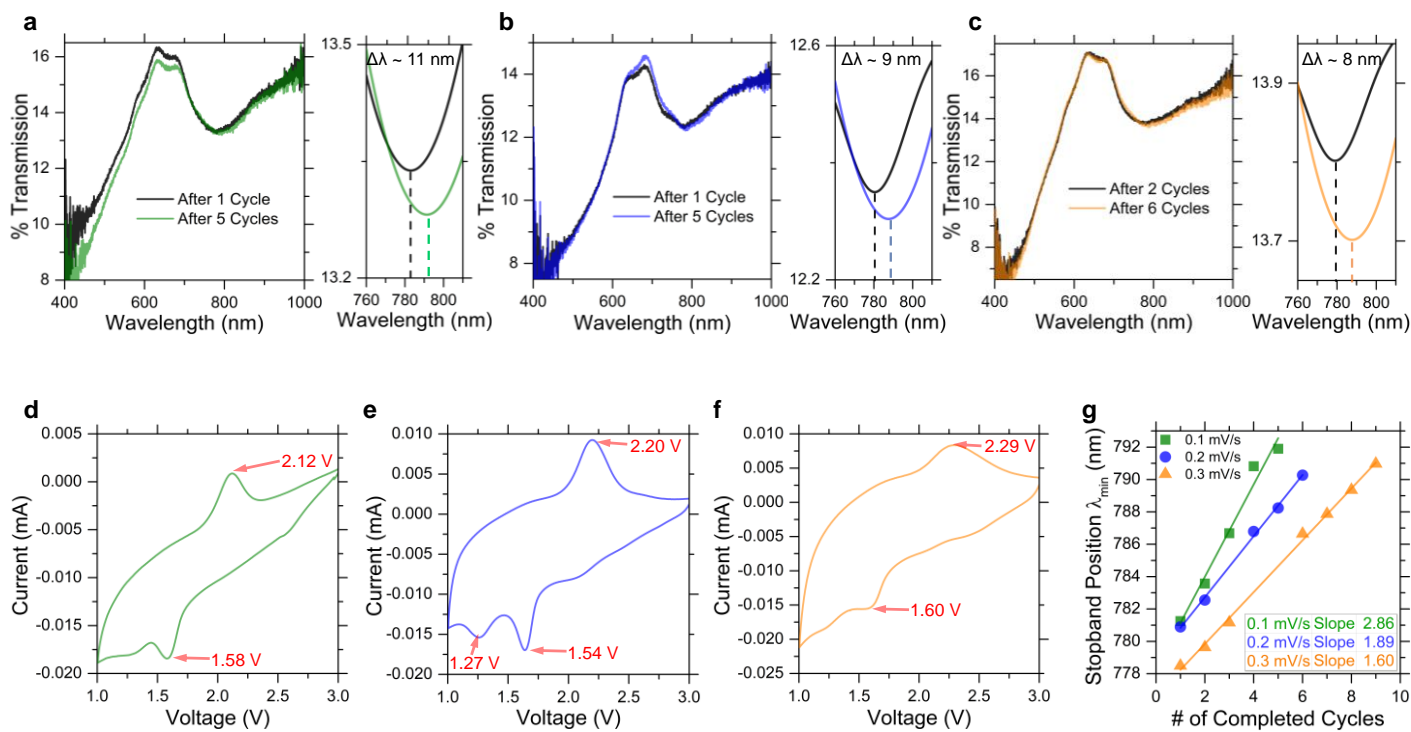

**Figure 4. Linking red-shift to electrode cycle life.** Gradual red-shifting of the photonic stopband in operando optical spectra of $TiO_2$ IOs occurring over the course of battery operation for cyclic voltammetry at rates of (a) 0.1 mV/s, (b) 0.2 mV/s and (c) 0.3 mV/s, all spectra shown are recorded at a voltage of 3.0 V. In each case, the shift in stopband position is shown with an accompanying plot where the shift in stopband position is highlighted. The corresponding cyclic voltammetry curves, with labelled reactions peaks, for the same $TiO_2$ IO anodes in their second cycle recorded at rates of (d) 0.1 mV/s (e) 0.2 mV/s and (f) 0.3 mV/s. (g) A plot of the photonic stopband positions, all recorded at 3.0 V, versus the number of completed CV cycles for $TiO_2$ IO anodes at three different CV scan rates. The data for the relationship between stopband position and number of completed cycles for fixed scan rate follow a linear fit.

An interesting feature of the plots in Figure 4 (g) is that there appears to be a strong linear correlation between the shift in stopband position and the number of cycles completed. This effect, for these electrodes with different CV scan speeds, is similar to the linear trend in stopband shift observed in galvanostatic charge/discharge, data displayed in Figure 3 (d). From the operando optical data, it would appear that every electrochemical test acts to alter the $TiO_2$ IO electrode, in a steady and predictable manner, giving rise to the observed photonic stopband red-shift effect. Additional tests performed on $TiO_2$ IOs prepared from larger polystyrene sphere templates also show the same optical effects; there is a red-shift in the photonic stopband (now centred at longer wavelengths from a larger initial pore size) with increased cycling. More details on the performance and behavior of the larger pore $TiO_2$ IO electrodes can be found in the Supplementary Materials Figs. S6-S8. The fact that the stopband red-shifts, regardless of the initial central wavelength position, suggests that this red-shift is closely linked to the properties of the photonic stopband and of the reversibly lithiated electrode as a whole and, critically, not an isolated or abnormal change to the transmission/absorption properties of the material in a specific region of the spectrum. Referring back to eqn. (1), there are two possible explanations for the red-shift in the photonic stopband: an increase in $d_{111}$ (the out-of-plane interplanar spacing of the photonic crystal) or an increase in $n_{\text{eff}}$ (the effective refractive index).

Considering the small magnitude of the red-shift to the photonic stopband, changes to either of these parameters would also be minor. In this study, some of the larger shifts in stopband position were on the scale of ~10 nm. In galvanostatic charge/discharge tests a 10 nm stopband red-shift was recorded after 70 cycles compared to just 7 cycles and in CV tests at 0.1 mV/s a stopband red-shift of 11 nm was recorded between cycle 1 and 5. The corresponding changes to the electrode parameters correspond $\Delta d_{111} = 3$ nm or $\Delta n_{\text{eff}} = 0.025$, assuming the optical shift can be attributed solely to each parameter and not some combination of both, overall constituting very small changes to the system properties. These miniscule modifications to the electrode system are difficult to detect accurately by experimental means, e.g. in situ or operando microscopy measurements require sensitivity to track the slight changes to $d_{111}$ which are of the order of ~1% of the overall interplanar spacing of the macroporous electrode, and photoelectron or energy loss spectroscopy analysis of the surface/interface rely on specific cell designs to confidently probe the region of interest. Tomographic or CT-methods could visualise structural changes, but the sensitivity of the spectral shifts used in this work provide high fidelity tracking of very small, but systematic, changes to the overall electrode during each cycle, and over the full cycle life, both together. Consequently, for the TiO$_2$ IO electrode system examined here, it is currently difficult to uncouple the minor shift in photonic stopband position and attribute its origin to either a change in $d_{111}$ or $n_{\text{eff}}$, or, some combination of both parameters.

The PhC electrodes examined in this work consisted of 8 – 10 layers of IO material (see Supplementary Materials Fig. S7, for example). Assuming a change in the structural periodicity of $\Delta d_{111} = 3$ nm alone is responsible for the photonic stopband shift, this would translate to a maximum increase in the overall height of the electrode of approximately 24 – 30 nm. Some expansion of the electrode height would indicate a change to $d_{111}$, otherwise a change to $n_{\text{eff}}$ could be assumed. Overall, the observations show a very sensitive change to optical transmission that is linked to the star of charge or depth of discharge (and lithium mole fraction) within each cycle. In addition, we can also track the small but systematic change to the periodicity and thickness of an electrode in real-time from the shift of photonic bandgap. Within each red-shifted cycle, the remarkably consistent reversible and cyclic change to transmission intensity during the cycle is always maintained and this is found under galvanostatic and voltammetric conditions. The operando optical effects observed here during reversible lithiation are fascinating and have the potential to offer an optical real-time readout of electrode state based on an observed structural colour. The method allow the reversibility to be tracked during each cycle and the state of the electrode over extended cycling non destructively, which may prove useful for material sensitive to swelling, changes in electronic conductivity or mechanical integrity with lithiation, or for assessing influence of SEI, CEI or other interfacial films or interphases.

**Conclusions**

Operando optical transmission spectra in conjunction with electrodes adopting an IO design have been shown to be capable of providing an optical assessment of the degree of lithiation and electrode cycling behaviour. Through close monitoring of the transmission spectra during operation, the operando optical spectra reveal an intricate relationship between the magnitude of the transmission intensity and the nominal lithium content in the electrode at various operating voltages. Drops in the overall transmission intensity correlate to higher uptake of lithium content in the electrode; a remarkable reversibility is seen in the inverse process as the intensity recovers at the onset of voltages corresponding to lithium extraction from the electrode. Cyclic variability in the optical transmission intensity is shown to relate directly to the state of charge of the electrode, linking the electrode optics to the electrochemical performance.

Another optical effect which shows great promise is the structural colour inherent to the IO PhC, exhibiting a direct response to repeated cycling. A gradual red-shift of the photonic stopband is incurred over the course of multiple cycles, establishing a visual colour indicator of the electrode behaviour and history. From the data sets reported here, it would appear that this shift in stopband position is linear with cycle number and, more importantly, predictable. An observable structural colour for an electrode linked to the cycle number has the potential to chronicle the electrochemical performance, provided the optical behaviour of the IO material is reasonably well-understood or characterised. Taking the broad applicability of this technique into consideration, with it being feasible for a wide range of electrode materials which can adopt an IO structure or be made to mimic one, the operando optical analysis explored here is versatile enough to inform on electrode phase and structural evolution while relating this information back to expected levels of lithiation during the cycle. This method shows great promise in assessing and diagnosing material performance, furthering our understanding of potential electrode candidates by closely monitoring their fundamental material properties alongside their electrochemical evolution.

## Acknowledgments

We acknowledge support from the Irish Research Council Government of Ireland Postgraduate Scholarship under award no. GOIPG/2016/946. We also acknowledge funding from the Irish Research Council Advanced Laureate Award under grant no. IRCLA/2019/118.

## Materials & Correspondence

Correspondence or request for materials should be address to Colm O'Dwyer, Email: c.odwyer@ucc.ie

# Methods

**Preparation of Polystyrene Artificial Opal Templates**

Polystyrene spheres in an aqueous suspension were used to form the sacrificial artificial opal templates for the IO structure. Monodisperse sphere suspensions, with a concentration of 2.5 wt% polystyrene, were purchased from Polysciences Inc. and used, as received, for forming the opal templates via dip-coating. In this work, spheres were purchased with diameters in the ranges of approximately 370 – 380 nm and 490 – 500 nm, depending on the application. A sulfate ester used in the preparation process of the polystyrene spheres bestowed a slight negative charge to the spheres. Opal templates were prepared on glass substrates which had been pre-cut into $2 \times 1$ cm$^2$ pieces. The glass substrates were first cleaned via successive ultrasonication (at a frequency of 40 kHz) in acetone (reagent grade 99.5%; Sigma-Aldrich), isopropyl alcohol (reagent grade 99.5%; Sigma-Aldrich) and deionised water, for 5 minutes each, in order to remove any dust or dirt particles by physical agitation. Following the cleaning process, a $1 \times 1$ cm$^2$ surface on the glass substrate was exposed to UV-Ozone treatment using a Novascan PSD Pro Series cleaner for a period of 1 hour, in order to promote surface hydrophilicity, immediately prior to immersion in the aqueous sphere suspension. Treated substrates were dip-coated into pre-heated vials of sphere suspension at a rate of 1 mm min$^{-1}$ which was maintained at a temperature of approximately 50 °C. An MTI PTL-MM01 dip-coater was used for all dip-coating processes. Glass substrates were immersed at a slight incline angle (~ 10 – 20°) to the vertical in order to improve the adhesion of the polystyrene spheres to the surface. Upon immersion into the suspension, the substrate was held still for 5 minutes to the allow the surface of the suspension to settle and allow a stable meniscus to form with the substrate. Finally, the substrate was withdrawn from the suspension at a rate of 1 mm min$^{-1}$ with the polystyrene spheres arranging themselves into an ordered colloidal crystal template during extraction.

**Sol-Gel Infiltration and Preparation of TiO$_2$ Inverse Opals**

TiO$_2$ inverse opals were prepared using the sol-gel synthesis technique to infiltrate the voids in a prepared sacrificial opal template. In this technique, a liquid (sol) precursor is introduced into the air voids of an opal template. Ambient moisture or applied heat to the infiltrated template allows the sol to hydrolyse, condense or crosslink to form a solid material (gel) which resides in the interstitial voids of the opal template. Annealing the samples at high temperatures is often performed to create crystalline materials. For the $TiO_2$ IOs prepared in this work, a 0.1 M $TiCl_4$ solution in isopropyl alcohol was utilised as the sol-gel precursor. A titanium (IV) chloride tetrahydrofuran complex ($TiCl_4 \cdot 2THF$ 97%; Sigma-Aldrich) was used as the source of $TiCl_4$; the precursor solution was magnetically stirred for 24 h before use, at which point the solution was clear in colour. For the sol-gel infiltration process, the $TiCl_4$ solution was drop-cast using a syringe onto opal templates. One or two drops of the precursor was applied, just enough to adequately wet the entire surface of the photonic crystal. Special care was taken to not flood the sample with excess precursor, which can create solid layers of material which may sit atop any potential IO structure. The sacrificial opal template was subsequently removed via calcination in air at 450 °C for 1 hr with a heating ramping rate of 5 °C $min^{-1}$. The high temperature allowed for both the removal of the opal template and for the crystallisation of the $TiO_2$ material to anatase phase $TiO_2$.

**Microscopy and Materials Characterisation**

Scanning electron microscopy (SEM) was carried out using a Zeiss Supra 40 high resolution SEM at typical accelerating voltages of 15–20 kV. The instrument was used to analyse the morphology of the opals and IO PhCs, and the metal film morphology or thickness. Dimensional analysis for feature size distributions from SEM images were measured using IMAGEJ software. Microscopy analysis was performed on selected samples before and after electrochemical cycling. Energy dispersive X-ray spectroscopy (EDX) analysis was also performed on the same SEM instrument using an accelerating voltage of 15 kV. EDX line scan measurements were obtained in several directions across the surface. EDX analysis was particularly useful for detecting the presence of residual electrolyte salt in post-cycling images, often presenting as enlarged IO walls. EDX analysis could screen for elements present in the electrolyte salt and ensure that post-cycling images were free from any salt contamination, please see Supplementary Materials Figs. S11 and S12 for further details.

**Raman Scattering Spectroscopy for Identifying Material Phase**

Raman scattering analysis was used to characterise the prepared $TiO_2$ IO films. The analysis was useful for identifying the material phase of $TiO_2$, with modes corresponding to anatase phase $TiO_2$ detected here (see

Supplementary Materials Fig. S10 for more details). In this work, Raman scattering spectroscopy was carried out using a Renishaw InVia Raman Spectrometer with a 30 mW Ar$^+$ laser with an excitation wavelength of 532 nm. The beam was focused using a 40x objective lens and collected using a RenCamCCD camera.

**Deposition of Conductive Metal Layers onto TiO$_2$ Inverse Opals**

Nickel metal layers were deposited over the surface of TiO$_2$ IOs, formed on glass substrates, in order to provide a conductive layer for the electrode. A thickness of 20 nm of Ni was used for the electrodes in this work, providing a good balance between the conductivity and the optical transmission of the electrode. The metal films were deposited via physical vapour deposition using a Quorum 150 T S magnetron sputtering system. The thickness of the film deposited could be measured using a built-in film thickness monitor. The Ni metal target was purchased from Ted Pella Inc. featuring a material purity of 99.99%. The magnetron sputtering system featured a shutter system which could be used to clean the surface of the Ni target prior to deposition, a necessary precaution in the case of Ni metal which is prone to oxidation. All films were sputtered under an inert atmosphere of high purity to prevent the any reaction between the metal particles and the gas atmosphere. In this case, argon gas (> 99.995%) was used as the inert gas for sputtering. The possibility of using transparent conductive oxides, such as fluorine-doped tin oxide, as substrates was explored and found to be incompatible with this measurement system due to alloying lithiation-induced opacity, see Supplementary Materials Fig. S1 for more details.

**Creating the Transparent Battery Cell for Recording Operando Spectra**

Conventional coin and pouch cells used in traditional and non-specialised lithium-ion battery research could not be used for analysis in this work as they do not possess the capability to record optical spectra, often been constructed and encapsulated in stainless steel. In order to record optical spectra consistently and to ensure compatibility with our optical system, a custom battery cell was designed to facilitate optical measurements. A cylindrical glass vial with a diameter of 2.5 cm and a height of 5 cm was used as the main body of the transparent cell, allowing for light to pass through and be collected by a spectrometer. A custom 3D printed cap was designed using AutoCAD to tightly to fit over the top of the system, while also gripping the electrode and ensuring it remained upright during analysis. Two protruding holes from the top of the cap were used to allow for electrical wire connections to the electrodes. A stereolithography (SLA) vat polymerisation 3D printer was used for constructing 3D models. Here, a Formlabs Form 2 3D printer, featuring a 250 mW laser operating at 405 nm, was used. High temperature resistant resin from Formlabs was used, on merit of its strong structural composition and high deflection temperature >210°C, in order to resist possible degradation from electrolyte vapour during operation. A hot glue gun and Kapton tape (Bemis Company, Incorporated) were used to seal openings in the cell and keep it hermetically sealed. All cells were assembled in an inert atmosphere, in this case inside a glovebox in an argon environment.

**Optical Transmission Measurements**

Optical transmission spectra were recorded using an Ocean Optics Inc. USB2000 + VIS-NIR-ES UV-visible spectrometer which features an operational wavelength range of 350 – 1000 nm. An unpolarised tungsten-halogen lamp with an operational wavelength range of 400 – 2200 nm purchased from Thorlabs Inc. was used as the broadband light source for recording spectra. For spectra recorded in air, 100% transmittance was normalised to the transmission spectrum of a clean blank piece of glass, allowing the spectra of $TiO_2$ IOs on glass to be recorded relative to the spectrum of glass. Similarly, for the spectra recorded in 0.1 M $LiPF_6$ electrolyte, 100% transmittance was normalised to the transmission spectrum of a clean blank piece of glass suspended upright in the electrolyte. For all spectra recorded in this work, the angle of incidence between the normal from the photonic crystal surface and the incident light was fixed at 0° incidence. For recording measurements over long periods of time (e.g. operando spectra recorded over many different electrochemical cycles), transparent battery cells were clamped in place to prevent any movement of the sample. This is an important consideration as there is some spectral variance experienced between different sample spots on the same IO electrode; clamping the cell in place ensured that the spectra were consistent and, importantly, recorded from the exact same sample spot.

**Electrochemical Analysis**

Electrochemical testing was carried out using a Biologic VSP potentiostat/galvanostat instrument. $TiO_2$ IO electrodes were tested in a half-cell arrangement, as the working electrode, with lithium metal acting as the counter and reference electrode. The mass loadings for the $TiO_2$ IO electrodes were within the range 0.1 – 0.2 mg. Figure 1 (b) shows the typical half-cell configuration used in this work, with the half-cell setup tailored for optical transmission. A flooded-cell design, with an excess of electrolyte of approximately 5 – 6 mL, was used for all of the electrochemical tests in this work. The electrolyte was purchased from Sigma-Aldrich as a 1 M solution of $LiPF_6$ salt dissolved in a 1:1 (v/v) mixture of ethylene carbonate and diethyl carbonate. For both galvanostatic and cyclic voltammetry tests, the voltage ranges were set to 1.0 – 3.0 V. CV scan rates were varied from 0.1 – 0.3 mV/s, depending on the application. For galvanostatic tests, C-rates were calculated based on a specific capacity of 168 mA h $g^{-1}$ for $TiO_2$, where 1 C corresponds to an applied specific current of 168 mA $g^{-1}$. In this work, galvanostatic tests were either carried out at 1 or 2 C, explicitly stated in each case.

# References

1. Kim, T., Song, W., Son, D.-Y., Ono, L. K. & Qi, Y. Lithium-ion batteries: outlook on present, future, and hybridized technologies. *Journal of Materials Chemistry A* **7**, 2942-2964 (2019). https://doi.org/10.1039/c8ta10513h
2. Wu, F., Maier, J. & Yu, Y. Guidelines and trends for next-generation rechargeable lithium and lithium-ion batteries. *Chemical Society Reviews* **49**, 1569-1614 (2020). https://doi.org/10.1039/c7cs00863e
3. Ziegler, M. S. & Trancik, J. E. Re-examining rates of lithium-ion battery technology improvement and cost decline. *Energy & Environmental Science* **14**, 1635-1651 (2021). https://doi.org/10.1039/d0ee02681f
4. Chen, D., Lou, Z., Jiang, K. & Shen, G. Device Configurations and Future Prospects of Flexible/Stretchable Lithium-Ion Batteries. *Advanced Functional Materials* **28**, 1805596 (2018). https://doi.org:https://doi.org/10.1002/adfm.201805596
5. Kong, L., Tang, C., Peng, H.-J., Huang, J.-Q. & Zhang, Q. Advanced energy materials for flexible batteries in energy storage: A review. *SmartMat* **1** (2020). https://doi.org:https://doi.org/10.1002/smm2.1007
6. Kim, H. *et al.* Exploring Anomalous Charge Storage in Anode Materials for Next-Generation Li Rechargeable Batteries. *Chemical Reviews* **120**, 6934-6976 (2020). https://doi.org/10.1021/acs.chemrev.9b00618
7. Wang, F. *et al.* Prelithiation: A Crucial Strategy for Boosting the Practical Application of Next-Generation Lithium Ion Battery. *ACS Nano* **15**, 2197-2218 (2021). https://doi.org/10.1021/acsnano.0c10664
8. Shi, J.-L. *et al.* High-Capacity Cathode Material with High Voltage for Li-Ion Batteries. *Advanced Materials* **30**, 1705575 (2018). https://doi.org:https://doi.org/10.1002/adma.201705575
9. Manthiram, A. A reflection on lithium-ion battery cathode chemistry. *Nature Communications* **11**, 1550 (2020). https://doi.org/10.1038/s41467-020-15355-0
10. Mahmood, N., Tang, T. & Hou, Y. Nanostructured Anode Materials for Lithium Ion Batteries: Progress, Challenge and Perspective. *Advanced Energy Materials* **6**, 1600374 (2016). https://doi.org:https://doi.org/10.1002/aenm.201600374
11. Zhang, L. *et al.* Advanced Matrixes for Binder-Free Nanostructured Electrodes in Lithium-Ion Batteries. *Advanced Materials* **32**, 1908445 (2020). https://doi.org:https://doi.org/10.1002/adma.201908445
12. Osiak, M., Geaney, H., Armstrong, E. & O'Dwyer, C. Structuring materials for lithium-ion batteries: advancements in nanomaterial structure, composition, and defined assembly on cell performance. *Journal of Materials Chemistry A* **2**, 9433-9460 (2014). https://doi.org/10.1039/c4ta00534a
13. Pomerantseva, E., Bonaccorso, F., Feng, X., Cui, Y. & Gogotsi, Y. Energy storage: The future enabled by nanomaterials. *Science* **366**, eaan8285 (2019). https://doi.org/10.1126/science.aan8285
14. Xu, J., Deshpande, R. D., Pan, J., Cheng, Y.-T. & Battaglia, V. S. Electrode Side Reactions, Capacity Loss and Mechanical Degradation in Lithium-Ion Batteries. *Journal of The Electrochemical Society* **162**, A2026-A2035 (2015). https://doi.org/10.1149/2.0291510jes
15. Liu, H. *et al.* Intergranular Cracking as a Major Cause of Long-Term Capacity Fading of Layered Cathodes. *Nano Letters* **17**, 3452-3457 (2017). https://doi.org/10.1021/acs.nanolett.7b00379
16. Iqbal, N., Haq, I. U. & Lee, S. Chemo-mechanical model predicted critical SOCs for the mechanical stability of electrode materials in lithium-ion batteries. *International Journal of Mechanical Sciences* **216**, 107034 (2022). https://doi.org:https://doi.org/10.1016/j.ijmecsci.2021.107034
17. Liu, D. *et al.* Review of Recent Development of In Situ/Operando Characterization Techniques for Lithium Battery Research. *Advanced Materials* **31**, 1806620 (2019). https://doi.org:https://doi.org/10.1002/adma.201806620
18. Li, H., Guo, S. & Zhou, H. In-situ/operando characterization techniques in lithium-ion batteries and beyond. *Journal of Energy Chemistry* **59**, 191-211 (2021). https://doi.org:https://doi.org/10.1016/j.jechem.2020.11.020
19. Grant, A. & O'Dwyer, C. Real-time nondestructive methods for examining battery electrode materials. *Applied Physics Reviews* **10**, 011312 (2023). https://doi.org/10.1063/5.0107386
20. Harks, P. P. R. M. L., Mulder, F. M. & Notten, P. H. L. In situ methods for Li-ion battery research: A review of recent developments. *Journal of Power Sources* **288**, 92-105 (2015). https://doi.org:https://doi.org/10.1016/j.jpowsour.2015.04.084
21. van de Krol, R., Goossens, A. & Meulenkamp, E. A. In Situ X-Ray Diffraction of Lithium Intercalation in Nanostructured and Thin Film Anatase TiO2. *Journal of The Electrochemical Society* **146**, 3150-3154 (1999). https://doi.org/10.1149/1.1392447
22. Misra, S. *et al.* In Situ X-ray Diffraction Studies of (De)lithiation Mechanism in Silicon Nanowire Anodes. *ACS Nano* **6**, 5465-5473 (2012). https://doi.org/10.1021/nn301339g

| | |
|---|---|
| 23 | Li, K. *et al.* Ultrafast-Charging and Long-Life Li-Ion Battery Anodes of TiO2-B and Anatase Dual-Phase Nanowires. *ACS Applied Materials & Interfaces* **9**, 35917-35926 (2017). https://doi.org/10.1021/acsami.7b11652 |
| 24 | Long, B. R., Chan, M. K. Y., Greeley, J. P. & Gewirth, A. A. Dopant Modulated Li Insertion in Si for Battery Anodes: Theory and Experiment. *The Journal of Physical Chemistry C* **115**, 18916-18921 (2011). https://doi.org/10.1021/jp2060602 |
| 25 | Kitta, M. & Sano, H. Real-Time Observation of Li Deposition on a Li Electrode with Operand Atomic Force Microscopy and Surface Mechanical Imaging. *Langmuir* **33**, 1861-1866 (2017). https://doi.org/10.1021/acs.langmuir.6b04651 |
| 26 | McAllister, Q. P., Strawhecker, K. E., Becker, C. R. & Lundgren, C. A. In situ atomic force microscopy nanoindentation of lithiated silicon nanopillars for lithium ion batteries. *Journal of Power Sources* **257**, 380-387 (2014). https://doi.org:https://doi.org/10.1016/j.jpowsour.2014.01.077 |
| 27 | Trease, N. M., Zhou, L., Chang, H. J., Zhu, B. Y. & Grey, C. P. In situ NMR of lithium ion batteries: Bulk susceptibility effects and practical considerations. *Solid State Nuclear Magnetic Resonance* **42**, 62-70 (2012). https://doi.org:https://doi.org/10.1016/j.ssnmr.2012.01.004 |
| 28 | Hovington, P. *et al.* In situ Scanning electron microscope study and microstructural evolution of nano silicon anode for high energy Li-ion batteries. *Journal of Power Sources* **248**, 457-464 (2014). https://doi.org:https://doi.org/10.1016/j.jpowsour.2013.09.069 |
| 29 | Fan, Z. *et al.* In Situ Transmission Electron Microscopy for Energy Materials and Devices. *Advanced Materials* **31**, 1900608 (2019). https://doi.org:https://doi.org/10.1002/adma.201900608 |
| 30 | Shi, F., Ross, P. N., Somorjai, G. A. & Komvopoulos, K. The Chemistry of Electrolyte Reduction on Silicon Electrodes Revealed by in Situ ATR-FTIR Spectroscopy. *The Journal of Physical Chemistry C* **121**, 14476-14483 (2017). https://doi.org:10.1021/acs.jpcc.7b04132 |
| 31 | Hsieh, A. G. *et al.* Electrochemical-acoustic time of flight: in operando correlation of physical dynamics with battery charge and health. *Energy & Environmental Science* **8**, 1569-1577 (2015). https://doi.org:10.1039/c5ee00111k |
| 32 | Merryweather, A. J., Schnedermann, C., Jacquet, Q., Grey, C. P. & Rao, A. Operando optical tracking of single-particle ion dynamics in batteries. *Nature* **594**, 522-528 (2021). https://doi.org:10.1038/s41586-021-03584-2 |
| 33 | Merryweather, A. J. *et al.* Operando monitoring of single-particle kinetic state-of-charge heterogeneities and cracking in high-rate Li-ion anodes. *Nature Materials* (2022). https://doi.org:10.1038/s41563-022-01324-z |
| 34 | Albero Blanquer, L. *et al.* Optical sensors for operando stress monitoring in lithium-based batteries containing solid-state or liquid electrolytes. *Nature Communications* **13**, 1153 (2022). https://doi.org:10.1038/s41467-022-28792-w |
| 35 | Hogrefe, C. *et al.* Cross-Sectional In Situ Optical Microscopy with Simultaneous Electrochemical Measurements for Lithium-Ion Full Cells. *Journal of The Electrochemical Society* **169**, 050519 (2022). https://doi.org:10.1149/1945-7111/ac6c57 |
| 36 | Zhang, D., Wang, R., Wang, X. & Gogotsi, Y. In situ monitoring redox processes in energy storage using UV–Vis spectroscopy. *Nature Energy* (2023). https://doi.org:10.1038/s41560-023-01240-9 |
| 37 | Sakamoto, J. S. & Dunn, B. Hierarchical battery electrodes based on inverted opal structures. *Journal of Materials Chemistry* **12**, 2859-2861 (2002). https://doi.org:10.1039/b205634h |
| 38 | McNulty, D., Lonergan, A., O'Hanlon, S. & O'Dwyer, C. 3D open-worked inverse opal TiO2 and GeO2 materials for long life, high capacity Li-ion battery anodes. *Solid State Ionics* **314**, 195-203 (2018). https://doi.org:https://doi.org/10.1016/j.ssi.2017.10.008 |
| 39 | McNulty, D., Carroll, E. & O'Dwyer, C. Rutile TiO2 Inverse Opal Anodes for Li-Ion Batteries with Long Cycle Life, High-Rate Capability, and High Structural Stability. *Advanced Energy Materials* **7**, 1602291 (2017). https://doi.org:https://doi.org/10.1002/aenm.201602291 |
| 40 | Armstrong, E. & O'Dwyer, C. Artificial opal photonic crystals and inverse opal structures – fundamentals and applications from optics to energy storage. *Journal of Materials Chemistry C* **3**, 6109-6143 (2015). https://doi.org:10.1039/c5tc01083g |
| 41 | Lonergan, A. & O'Dwyer, C. Many Facets of Photonic Crystals: From Optics and Sensors to Energy Storage and Photocatalysis. *Advanced Materials Technologies*, 2201410 (2022). https://doi.org:https://doi.org/10.1002/admt.202201410 |
| 42 | Schroden, R. C., Al-Daous, M., Blanford, C. F. & Stein, A. Optical Properties of Inverse Opal Photonic Crystals. *Chemistry of Materials* **14**, 3305-3315 (2002). https://doi.org:10.1021/cm020100z |
| 43 | Lonergan, A., McNulty, D. & O'Dwyer, C. Tetrahedral framework of inverse opal photonic crystals defines the optical response and photonic band gap. *Journal of Applied Physics* **124**, 095106 (2018). https://doi.org:10.1063/1.5033367 |

44  Zuo, Y. *et al.* A photonic hydrogel for health self-monitoring of solid-state electrolytes in zinc-air batteries. *Energy Storage Materials* **53**, 136-147 (2022). https://doi.org:https://doi.org/10.1016/j.ensm.2022.08.047
45  O'Dwyer, C. Color-Coded Batteries – Electro-Photonic Inverse Opal Materials for Enhanced Electrochemical Energy Storage and Optically Encoded Diagnostics. *Advanced Materials* **28**, 5681-5688 (2016). https://doi.org:https://doi.org/10.1002/adma.201503973
46  Lonergan, A., Hu, C. & O'Dwyer, C. Filling in the gaps: The nature of light transmission through solvent-filled inverse opal photonic crystals. *Physical Review Materials* **4**, 065201 (2020). https://doi.org:10.1103/PhysRevMaterials.4.065201
47  Lonergan, A., Murphy, B. & O'Dwyer, C. Photonic Stopband Tuning in Metallo-Dielectric Photonic Crystals. *ECS Journal of Solid State Science and Technology* **10**, 085001 (2021). https://doi.org:10.1149/2162-8777/ac19c5
48  Gao, X. *et al.* Preparation and Electrochemical Characterization of Anatase Nanorods for Lithium-Inserting Electrode Material. *The Journal of Physical Chemistry B* **108**, 2868-2872 (2004). https://doi.org:10.1021/jp036821i
49  Laskova, B., Zukalova, M., Zukal, A., Bousa, M. & Kavan, L. Capacitive contribution to Li-storage in TiO2 (B) and TiO2 (anatase). *Journal of Power Sources* **246**, 103-109 (2014). https://doi.org:https://doi.org/10.1016/j.jpowsour.2013.07.073
50  Ou, J., Wu, S., Yang, L. & Wang, H. Facile Preparation of NiO@graphene Nanocomposite with Superior Performances as Anode for Li-ion Batteries. *Acta Metallurgica Sinica (English Letters)* **35**, 212-222 (2022). https://doi.org:10.1007/s40195-021-01283-5

# Operando Photonic Band Gap Probe of Battery Electrode Materials

Alex Lonergan[1], Umair Gulzar[1], Yan Zhang[1] and Colm O'Dwyer[1,2,3,4]*

[1]*School of Chemistry, University College Cork, Cork, T12 YN60, Ireland*
[2]*Micro-Nano Systems Centre, Tyndall National Institute, Lee Maltings, Cork, T12 R5CP, Ireland*
[3]*AMBER@CRANN, Trinity College Dublin, Dublin 2, Ireland*
[4]*Environmental Research Institute, University College Cork, Lee Road, Cork T23 XE10, Ireland*

## Contents

### 1. Optical opacity associated with transparent conductive oxide substrates

In the main text, we described how a custom electrode needed to be designed to circumvent issues with substrate transparency for conductive oxide glasses. The final electrode design uses a non-conductive glass substrate to support the active material (a $TiO_2$ inverse opal) with 20 nm of Ni metal deposited over the entire surface to act as a current collector layer. The thin layer of metal acting as the current collector layer is not ideal in terms of conductivity, yet thicker layers could not be used due to the need for optical transparency. Ideally, the conductive layer would be highly transparent to optical wavelengths. For this reason, glass substrates with fluorine-doped tin oxide (FTO) layers were first tested as potential conductive substrates for $TiO_2$ IOs. Figure S1 (a) depicts a typical $TiO_2$ IO prepared on an FTO substrate indicating the relative thickness of the layers comprising the electrode structure. In this instance, a $TiO_2$ IO with a larger periodicity to those depicted in the main text was used (see Figs S6 – S8 below for more details on larger IO periodicities), hence the red hue in the observed structural colour.

The optical transmission spectra for a $TiO_2$ IO, prepared on an FTO substrate, in an air and electrolyte background is shown in Fig. S1 (b). In air, a photonic stopband is observable with a central wavelength position of 693 nm. In electrolyte, the photonic stopband is red-shifted to longer wavelengths, with a central wavelength position of 892 nm. Figure S1 (c) contains several operando optical transmission spectra recorded during the first discharge of a $TiO_2$ IO prepared on an FTO substrate. A very noticeable effect which occurs for the FTO substrates is a significant reduction in optical transmittance for the electrode, occurring in the early stages of the battery life. The lithium insertion potential for $TiO_2$ occurs at ~ 1.65 V, as per the CV depicted in Figs. 2 (a) and (b) of the main text. The onset of the optical opacity for this electrode system occurs at higher potentials, suggesting it may be linked to a reaction between the $SnO_2$ in FTO and the lithium. The electrode visually darkens and creates an overall reduction in transparency.

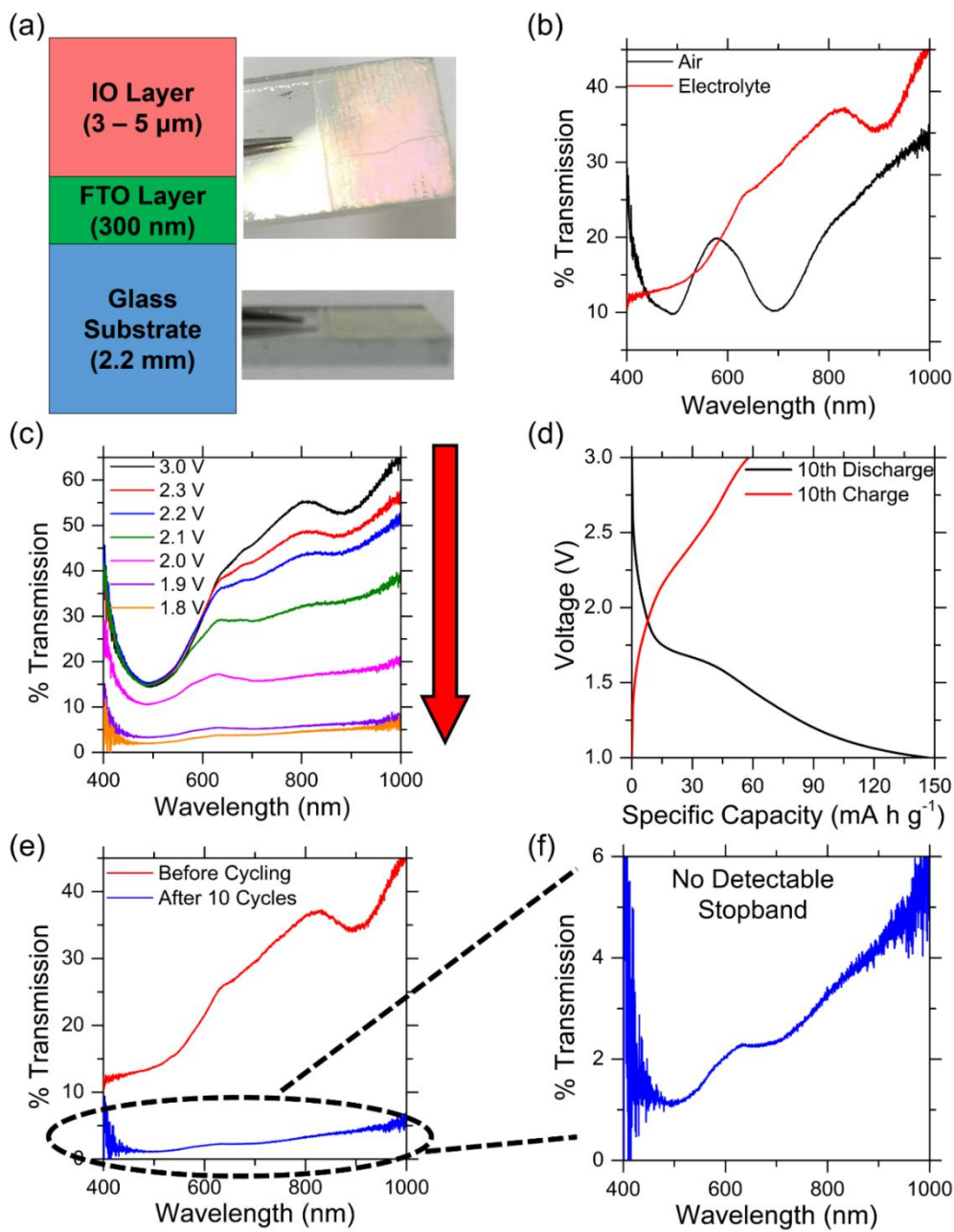

**Fig. S1** (a) A typical TiO$_2$ IO electrode prepared on an FTO substrate with a visible structural colour reflection. The relative thicknesses of the layers comprising the electrode are indicated. (b) Optical transmission spectra of a TiO$_2$ IO electrode prepared on an FTO substrate in an air and electrolyte background with prominent photonic stopbands present. (c) Operando transmission spectra recorded during the first discharge of a TiO$_2$ IO electrode prepared on an FTO substrate, showing a rapid decrease in overall transmittance. (d) Galvanostatic charge/discharge showing calculated specific capacities in the 10$^{th}$ cycle recorded at a specific current rate of 1C. (e) A comparison of the operando optical transmission spectra recorded before cycling and after completing 10 cycles. (f) A zoomed-in look at the operando optical spectrum after 10 cycles with no detectable stopband.

Electrochemically, the TiO$_2$ IO electrode on FTO exhibits a reasonably similar performance to the TiO$_2$ IO electrodes prepared on glass substrates with a conductive Ni layer which can be seen by comparing the calculated specific capacity values in Fig 4 (a) in the main text to Fig. S1 (d). However, while the electrochemical performance is similar in both cases, the data which can be extracted from the operando optical spectra is markedly different. Figure S1 (e) shows a comparison between the initial optical spectrum before any cycling occurs and the optical spectrum of the same IO sample on FTO after 10 cycles. Firstly, the overall transmittance of the electrode has significantly reduced, presumably linked to the darkening of the $SnO_2$ material in the conductive FTO coating, as discussed above. The optical features on the operando spectrum after 10 cycles can be more clearly seen in Fig. S1 (f). Unlike the operando spectra shown in the main text, there does not appear to be any detectable stopband present in the spectra for the $TiO_2$ IOs prepared on FTO. One possible explanation for the lack of a detectable photonic stopband could be that blackened $SnO_2$ layer in FTO, which most likely quickly developed via reaction of $SnO_2$ with lithium, obscured the presence of the photonic stopband in this case. Unfortunately, this means that it is not possible to track the evolution of the photonic stopband of the active material while use a conductive oxide substrate (like FTO), necessitating the need to use an alternative substrate, as outlined in the main text.

## 2. Additional SEM images of $TiO_2$ IO electrodes

An important consideration, for tracking the photonic stopband of the $TiO_2$ IO electrodes, is that the structural order of the IO is maintained during cycling. The photonic stopband is sensitive to the repeating lattice structure and the refractive index of the IO material. Hence, it is important to establish that the long range order and periodicity of the IO is retained after electrochemical cycling. Figure S2 (a) displays additional SEM images which show the condition of a $TiO_2$ IO which has performed 10 CV cycles at a rate of 0.3 mV/s, operando optical spectra and electrochemical for this IO can be found in the main text Figs. 4 (c) and (f), respectively. Likewise, Fig. S2 (b) shows additional SEM images for the a $TiO_2$ IO which has been cycled for 70 galvanostatic cycles at a rate of 2C, the corresponding electrochemical and operando optical spectra for the same sample can be found in the main text Fig. 3. The periodicity of the IO structure is still present in the all of the SEM images shown after undergoing electrochemical cycling. At higher magnifications, slight differences to the appearance of the top layer of IO material might arise from the rinsing of the electrode (to remove residual electrolyte salt) and re-coating with metal (to provide conductivity for microscopy imaging), see Fig. S9 below for more details on this process.

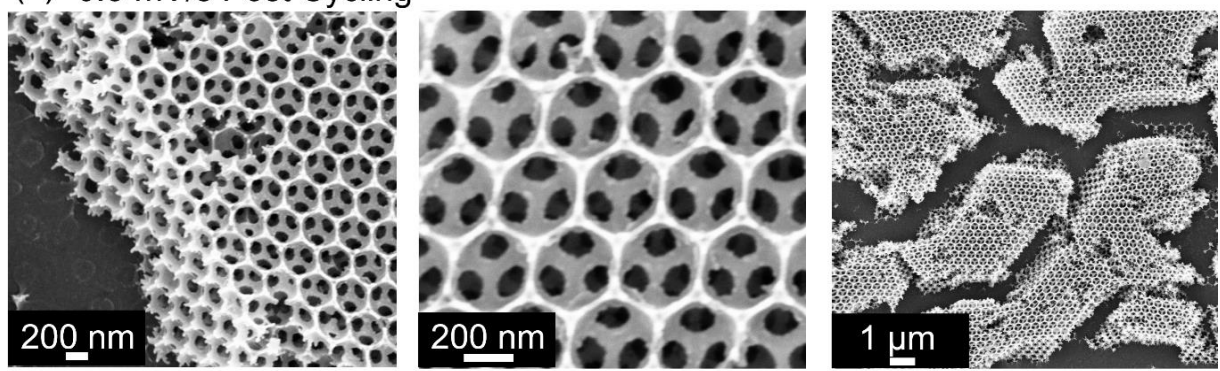

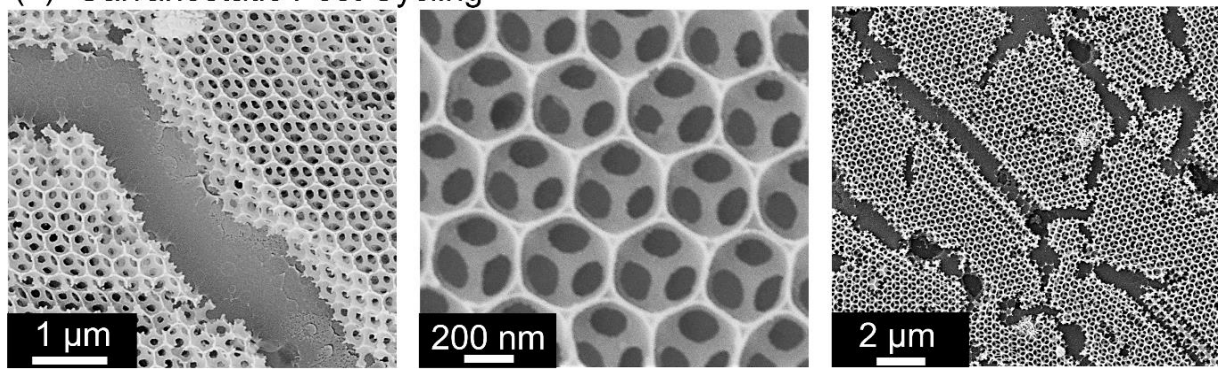

**Fig. S2** (a) SEM images of various magnifications showing the appearance of a $TiO_2$ IO electrode following 10 completed CV cycles at a rate of 0.3 mV/s. (b) SEM images of various magnifications showing the appearance of a $TiO_2$ IO electrode following 75 galvanostatic charge/discharge cycles at specific current rate of 2C.

## 3. Additional operando optical data and electrochemical characterisation

In the main text, a large drop in transmission intensity, occurring over the first electrochemical cycle only, was attributed to the formation of NiO from the Ni metal deposited over the surface to act as a conductive layer. Using a potential window of 3.0 – 1.0 V for testing the $TiO_2$ active material, the reconversion of NiO to Ni metal does not occur as this reaction potential is reported as occurring below 1.0 V[1]. This restricts the conversion reaction of Ni to NiO to the first discharge only, linking this reaction to the drop in overall transmission intensity after the first cycle. This effect was discussed in some detail in the main text. However, an optical effect which was only briefly mentioned was the effect of this supposed conversion of Ni metal to NiO on the position of the photonic stopband. Referring to eqns. (1) and (2) in the main text, the wavelength position of the photonic stopband is linked to the effective refractive index of the IO. The effective refractive index is dependent on the material comprising the IO network, including any material which exist in the interstitial voids of the structure. Although only a thin coating of Ni metal is deposited across the surface of the IO, constituting only a very small fraction of the overall material of the IO network, the conversion of Ni metal to NiO (and the subsequent increase in refractive index) should have an impact on the position of the photonic stopband[2].

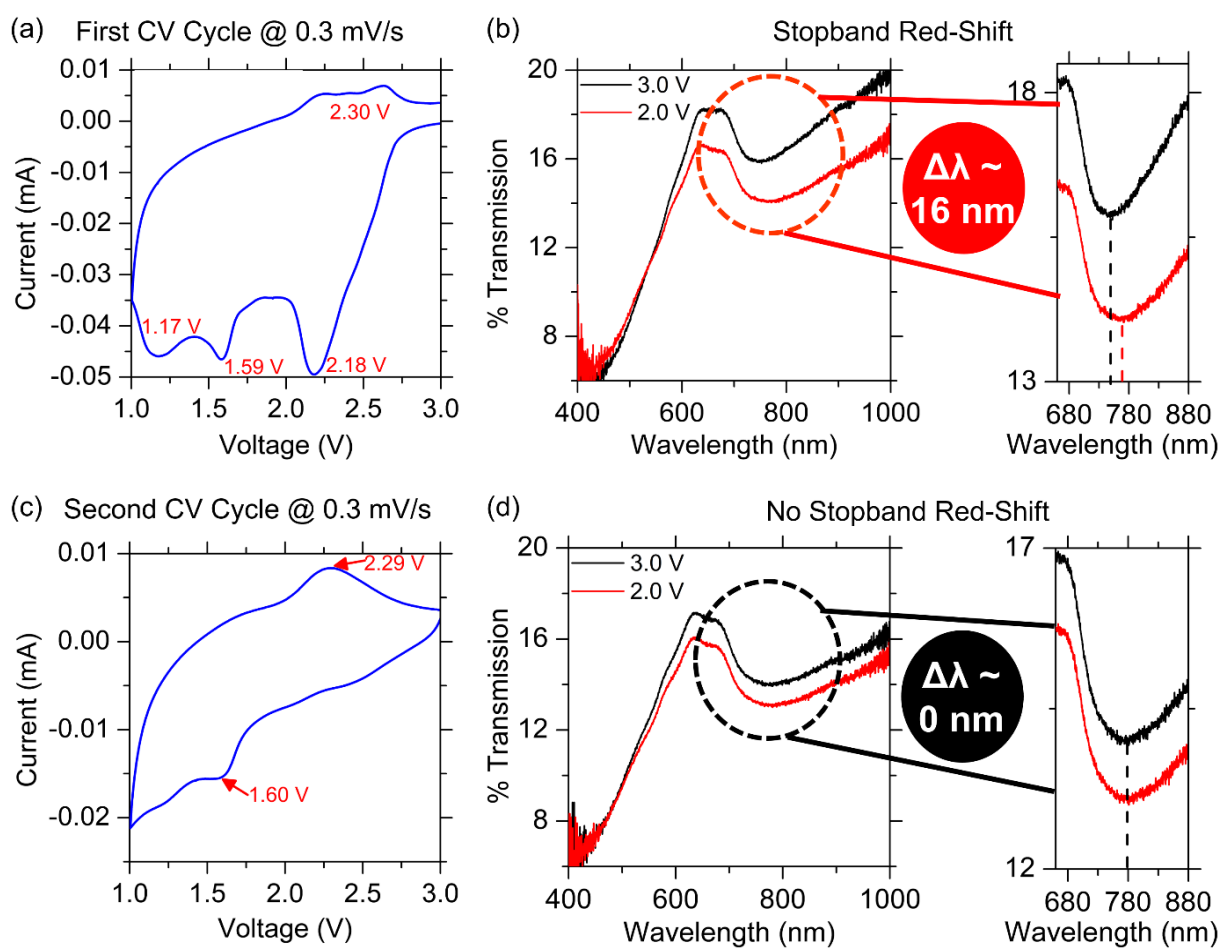

**Fig. S3** (a) Cyclic voltammetry profile with labelled reaction peaks for the first cycle of a $TiO_2$ IO electrode cycled at a rate of 0.3 mV/s. (b) Operando optical spectra recorded at 3.0 V and 2.0 V in the first cathodic sweep in CV testing. An enlarged optical spectrum highlights a large initial red-shift in stopband position of 16 nm occurring in the discharge from 3.0 – 2.0 V. (c) Cyclic voltammetry profile with labelled reaction peaks for the second cycle of the same $TiO_2$ IO electrode cycled at a rate of 0.3 mV/s. (d) Operando optical spectra recorded at 3.0 V and 2.0 V in the second cathodic sweep in CV testing. No detectable stopband red-shift occurs in the second CV cycle in the discharge from 3.0 V – 2.0 V.

Figure S3 (a) shows the first CV cycle for a TiO$_2$ IO electrode cycled at a rate of 0.3 mV/s. The prominent reaction at 2.18 V in the first cathodic sweep was linked to the reaction for the conversion of Ni metal to NiO in the main text. Figure S3 (b) displays the operando optical spectra recorded during the first cathodic sweep at 3.0 V and 2.0 V. At 2.0 V there is a stopband red-shift of approximately 16 nm from the initial position prior to cycling, a significant red-shift for this stage in the battery life. Importantly, this red-shift at 2.0 V is detected before the expected lithium insertion reaction for TiO$_2$ (~ 1.59 V, here), linking it to the conversion of Ni metal to NiO. In the second CV cycle for the same TiO$_2$ IO sample, shown in Fig. S3 (c), the prominent reaction at 2.18 V present in the first cathodic sweep is not detected. This supports the assertion that the for the potential window operated here, that is 3.0 – 1.0 V, the conversion of Ni metal to NiO is an irreversible reaction for this system. Figure S3 (d) shows the operando optical spectra recorded in the second CV cycle at 3.0 V and 2.0 V. In this case, there is no detectable red-shift in stopband position at 2.0 V. This suggests that the large initial red-shift after the first cycle is linked to the conversion of Ni metal to NiO.

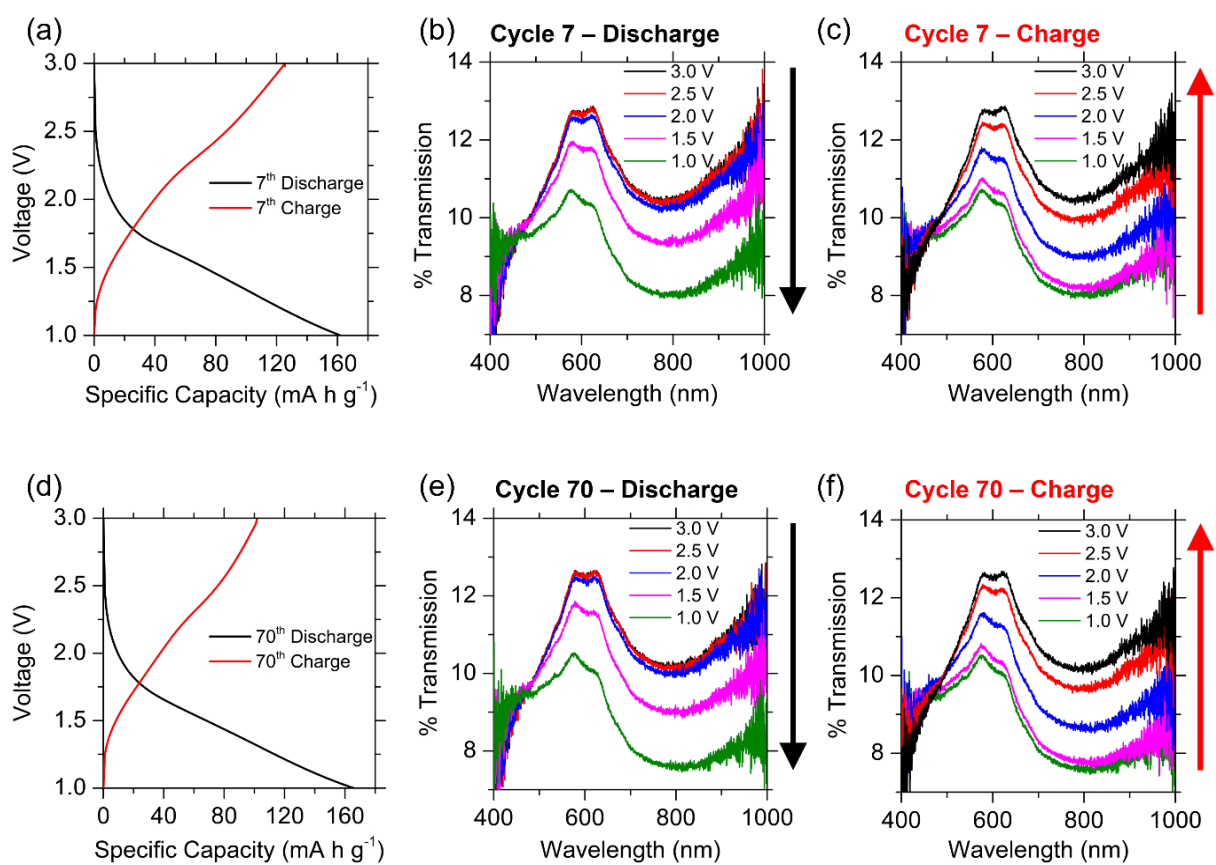

**Fig. S4** Calculated specific capacities for (a) cycle 7 and (d) cycle 70 for a TiO$_2$ IO electrode cycled at a rate of 2C. The accompanying operando optical transmission spectra recorded in 0.5 V intervals for the (b) cycle 7 discharge, (c) cycle 7 charge, (e) cycle 70 discharge and (f) cycle 70 charge. The trends in the fluctuation of optical transmittance with operating voltage during the discharge and charge processes are consistently observed in both cycle 7 and cycle 70.

Figure S4 provides additional electrochemical and optical data for a TiO$_2$ IO which performed galvanostatic charge/discharge cycling at a rate of 2C to accompany the data shown in Fig. 3 in the main text. The calculated specific capacities for the charge and discharge processes for cycle 7 and cycle 70 are displayed in Figs. S4 (a) and (d), respectively. More detailed discussion of the electrochemical performance of the TiO$_2$ IO can be found in the main text. Of particular note in Fig. S4 are the operando optical spectra. Figure 3 in the main text detailed the photonic stopband red-shifting effect which occurs with increased cycling. Here, the operando optical spectra in Fig. S4 illustrate how the cyclic changes in transmission intensity (reported in Fig. 2 in the main text for more discussion) are always occurring throughout the battery life for each cycle.

Figures S4 (b) an (e) show operando optical spectra in 0.5 V intervals for the discharge processes in cycle 7 and cycle 70, respectively. As discussed in the main text, increasing lithiation of the $TiO_2$ IO electrode acts to the decrease the overall transmission intensity; the largest drops in the intensity are closely linked to the voltages associated with lithiation insertion into $TiO_2$. In other words, electrode transmittance decreases with the nominal lithium content expected to be incorporated into the electrode material at a particular voltage. Likewise, Figs S4 (c) and (f) show the operando optical spectra in 0.5 V intervals for the charge processes in cycle 7 and 70, respectively. As lithium is removed from the $TiO_2$ IO electrode during the charge process, the overall optical transmittance of the increases. When fully charged (at 3.0 V), the appearance of the transmission spectrum has recovered to its initial state prior to the cycle beginning. This cyclic relationship in the operando transmission spectra is present for every cycle monitored throughout the battery life.

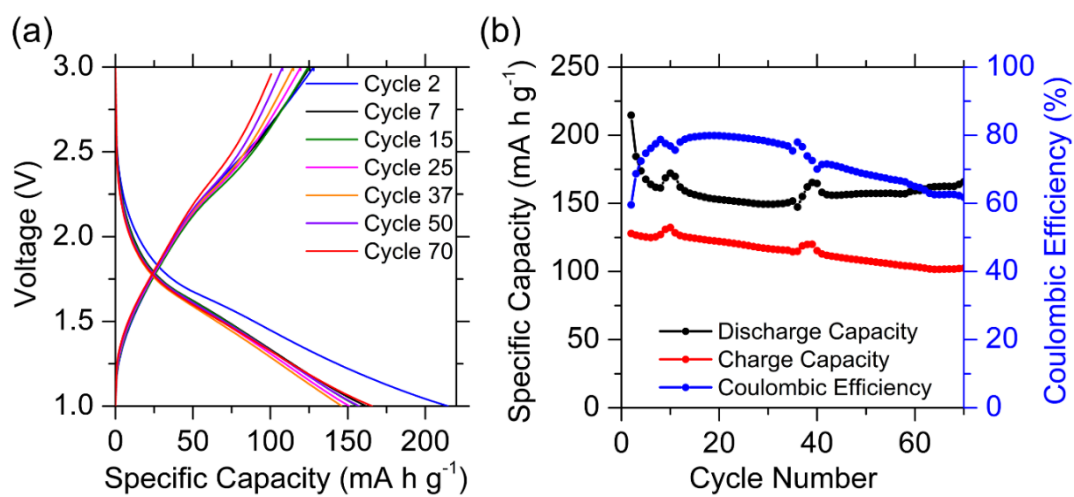

**Fig. S5** (a) Calculated specific capacities for the galvanostatic charge and discharge processes for a number of cycles of a typical $TiO_2$ IO cycled at a rate of 2C. (b) A plot showing the evolution of the specific capacity for the discharge and charge processes for the first 70 cycles of the same $TiO_2$ IO cycled at 2C. The calculated coulombic efficiency for the first 70 cycles is also shown.

Additional electrochemical data for the same $TiO_2$ IO electrode, cycled at a rate of 2C, is shown in Fig. S5. The calculated specific capacities for a number of cycles are shown for the discharge and charge processes in Fig. S5 (a). There is a noticeable difference in the discharge and charge specific capacity, with the charge process consistently featuring a smaller specific capacity. This effect occurs for every $TiO_2$ IO we tested with our custom set-up. In terms of the coulombic efficiency for the battery cell, Fig. S5 (b) shows the evolution of the coulombic efficiency for the cell, alongside the discharge and charge capacity, across the first 70 cycles in the battery life. The coulombic efficiency fluctuates between 60 – 80 %.

Many works with $TiO_2$ IO electrodes, prepared on stainless steel substrates and using traditional coin-cell set-ups, feature much higher coulombic efficiencies with values close to 100% [3][4]. Here, it is likely that the custom electrode design, needed to ensure optical transparency, limits the overall coulombic efficiency. For the custom set-up used in this work, a thin layer of Ni (20 nm) is used as the current collector layer. A thin layer of metal deposited over the surface of the entire electrode is not an ideal method to provide electrical conductivity, yet a thin layer was necessary to ensure optical transparency. Considering the differences between our custom electrode set-up and conventional battery testing methods in the literature, it is doubtless that the battery tested here features a much higher internal resistance. The coulombic efficiency of a cell is linked to number of factors, including: operating current, state of charge, the presence of parasitic reactions, temperature, battery age and internal resistance[5][6]. It is certainly plausible to attribute the likely higher internal resistances of the cells tested here as a contributing factor to the consistently observed lower coulombic efficiencies.

## 4. Larger IO periodicity: Equivalent optical effects and stopband shifting

The TiO$_2$ IOs used to investigate the operando optical effects in the main text were prepared using polystyrene sphere templates with diameters in the 370 – 380 nm range. In terms of the optical effects observed from the operando transmission spectra, a cyclic fluctuation of the electrode transmittance and a gradual red-shift of the photonic stopband position with cycling are observed for all TiO$_2$ IO electrodes tested. For the red-shifting of the photonic stopband position, it is important to establish that any changes to the appearance of the photonic stopband are linked to structural or refractive index changes in the ordered material and *not* occurring as a result of a change in the optics of the background electrolyte over time. Understanding how the structure or composition of the electrode evolves over time is the more salient effect which can occur here; monitoring changes in the photonic stopband of the electrode has the potential to establish a system which can use the optical signature of the electrode to diagnose and assess the performance of and the changes to the electrode, non-destructively and in-operando.

One way in which to determine whether the stopband red-shifting effect is tied to background optical effects or changes to the structural/optical properties of the IO, is to move the initial position of the photonic stopband and observe if the same optical red-shifting of the stopband occurs in different wavelength ranges. It is unlikely that changes to optics of the background electrolyte would homogenously span over large wavelength ranges, effectively eliminating this possibility if photonic stopband red-shifting effects were still observed. For this reason, larger TiO$_2$ IOs were prepared using polystyrene sphere templates with diameters in the 490 – 500 nm range. These larger TiO$_2$ IOs were prepared and used as electrodes in an identical fashion to the smaller IOs used throughout the main text. Figures S6 (a) and (b) show SEM images of the unmodified TiO$_2$ IO and the Ni-coated TiO$_2$ IO, respectively.

Figures S6 (c) and (d) show the appearance of the transmission spectra for the larger TiO$_2$ IOs in air and in electrolyte (also coated with 20 nm of nickel metal), respectively. In air, the photonic stopband for these IOs are centred at approximately 660 nm. After coating with 20 nm of Ni and immersing in electrolyte, the photonic stopband red-shifts to approximately 874 nm, as expected due to the increase in effective refractive index. These stopband positions constitute a significant shift in wavelength position from the IOs with smaller periodicities used in the main text (see Fig. 1) which featured a stopband at 573 nm in air and 766 nm in electrolyte. Optically, the effects observed in the operando transmission spectra for the larger TiO$_2$ IOs are identical to the effects observed for the IOs with a smaller periodicity in the main text. Figure S6 (e) illustrates that an irreversible drop in transmission intensity is observed after the first cycle only, matching the effects described by Figs. 3 (a) and (b) in the main text. Figure S6 (f) demonstrates that the stopband red-shifting effect is still observed in the case of theses IOs featuring a larger periodicity. Here, there is a red-shift of 11 nm which occurs between cycle 26 and 55 of a TiO$_2$ IO cycled at a rate of 1C. Observing that the photonic stopband red-shift still occurs at these longer wavelengths, lends more credibility to the assertion that the changes to the stopband are tied to structural/optical changes of the IO material and *not* related to background optical changes in the electrolyte spectral signature.

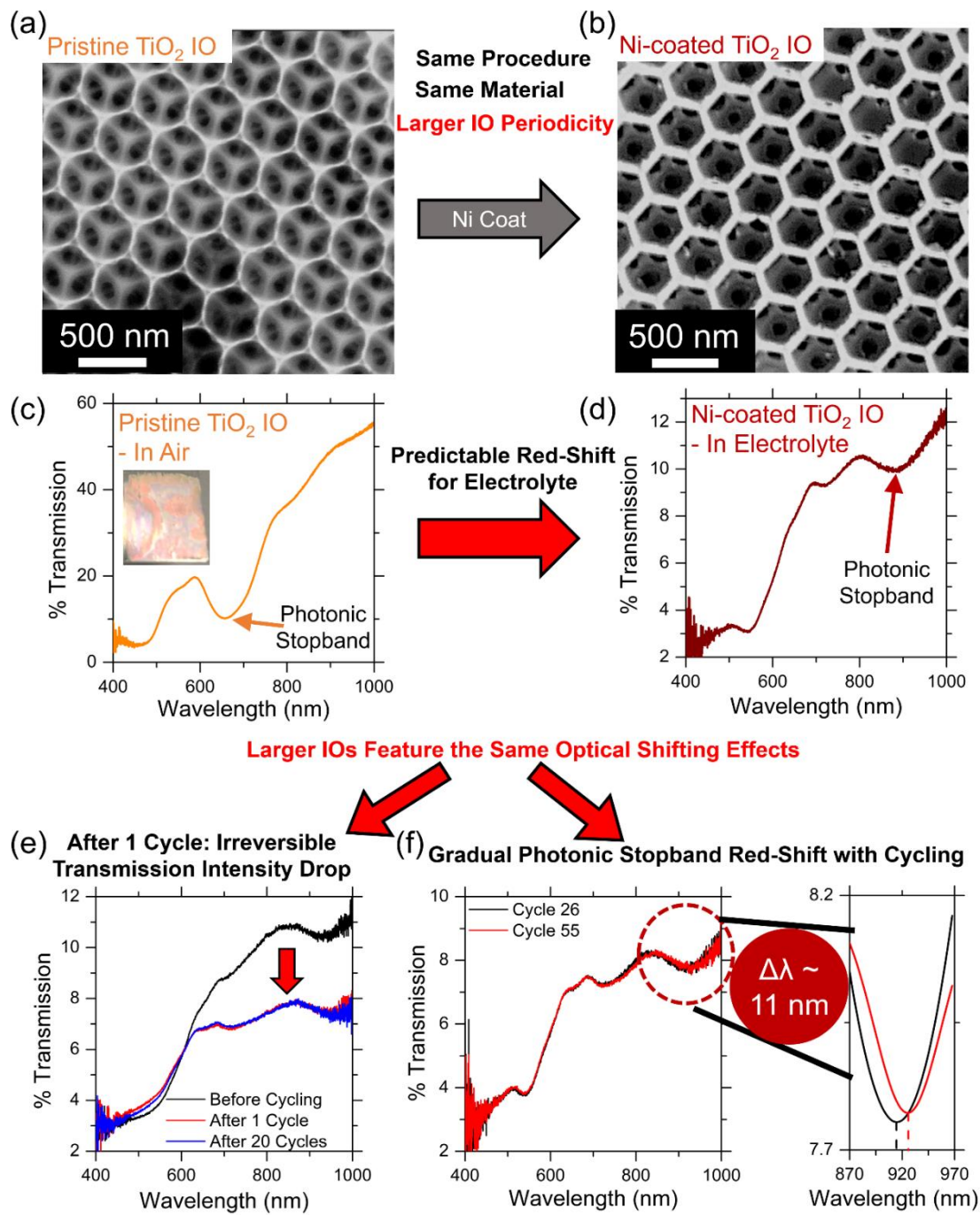

**Fig. S6** SEM images of a TiO$_2$ IO prepared using larger (500 nm) polystyrene opal templates showing the (a) pristine condition of the IO and (b) the IO after coating with 20 nm of Ni metal for electrical conductivity. The transmission spectra of a typical TiO$_2$ IO prepared from larger (500 nm) polystyrene opal templates with photonic stopband positions highlighted for IOs immersed (c) in air and (d) in electrolyte. TiO$_2$ IOs with larger periodicities show analogous optical effects to those presented in the main text through the operando optical spectra, namely the (e) irreversible transmission intensity drop after cycle 1 and (f) photonic stopband red-shifting with increasing cycle number.

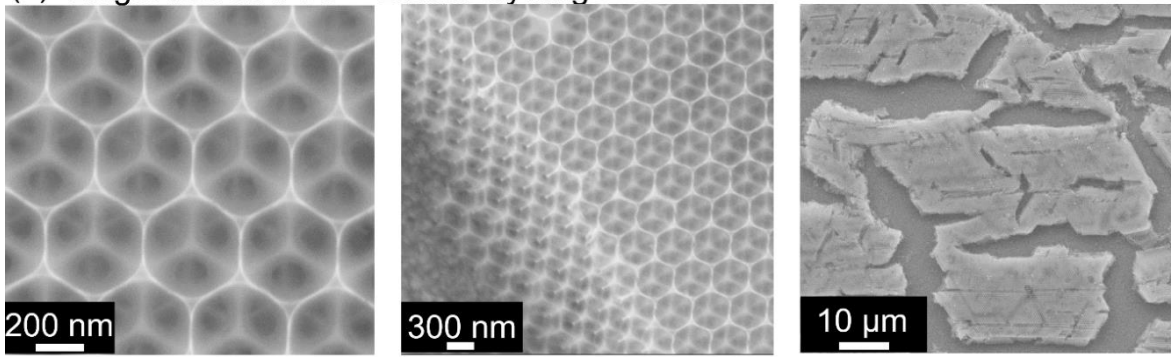
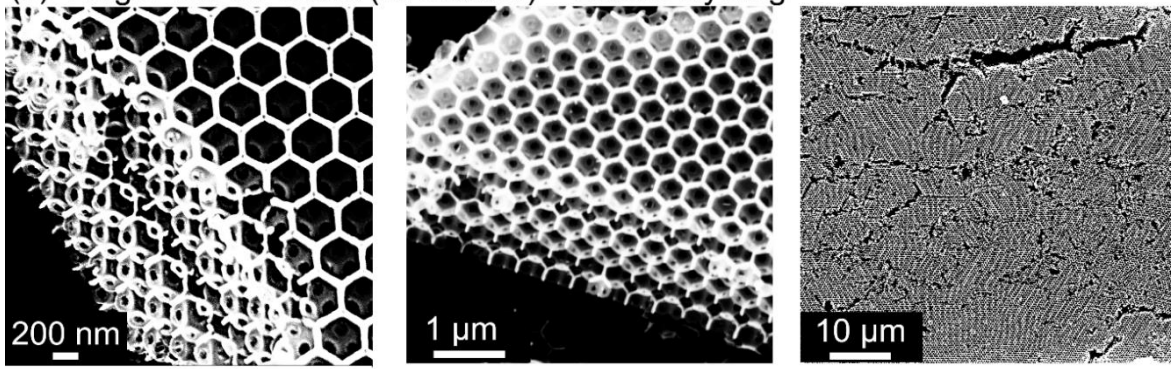
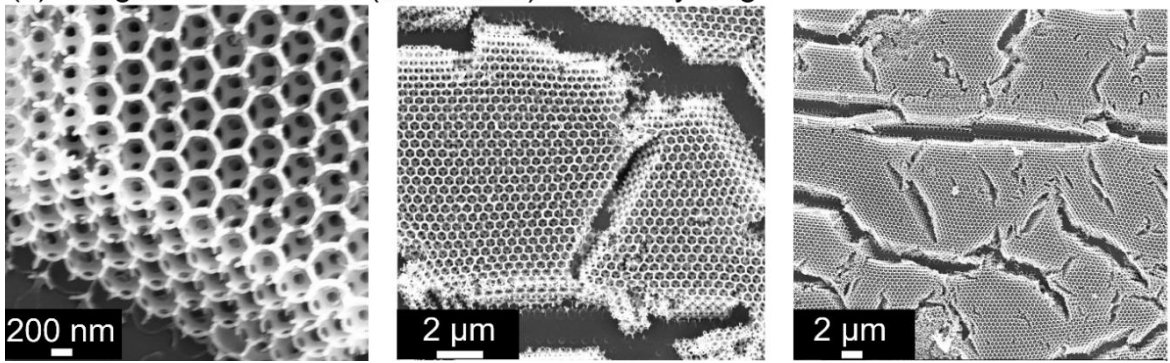

**Fig. S7** A collection of SEM images for TiO$_2$ IOs with larger periodicities prepared using 500 nm polystyrene opal templates. (a) SEM images of a TiO$_2$ IO prepared on an FTO substrate. (b) SEM images of a TiO$_2$ IO prepared on non-conductive glass and coated with a layer of 20 nm of Ni to act as a conductive layer. (c) SEM images of a TiO$_2$ IO which has been imaged after completing approximately 75 galvanostatic cycles at a rate of 1C.

Figure S7 contains a collection of SEM images characterising the TiO$_2$ IOs with larger periodicities used in this section. Figure S7 (a) shows the appearance of the TiO$_2$ IOs, prior to any electrochemical cycling, prepared on an FTO substrate. The appearance of the TiO$_2$ IOs, with thin IO walls, is easily observed here due the conductive layer of the FTO eliminating the need for a Ni metal conductive layer for imaging. Figure S7 (b) displays SEM images for larger TiO$_2$ IOs prepared on a non-conductive glass substrate with 20 nm on Ni deposited to act as a conductive later for imaging. Lastly, Fig. S7 (c) shows that appearance of these larger TiO$_2$ IOs, prepared on glass substrates, post-cycling at a rate of 1C. Importantly, the structure of the IO is largely retained after cycling. In all cases shown in Fig. S7, there appears to be approximately 8 – 10 layers of IO material in the layered structure with large 'islands' of IO material present throughout the structure.

One final observation on the optical behaviour of these larger TiO$_2$ IOs is illustrated in Fig. S8. Much like in the case of the TiO$_2$ IOs with smaller periodicities explored in the main text, the overall transmittance of the electrode obeys a cyclic behavioural pattern where transmittance is highest for nominally low lithium content incorporated into the electrode and vice-versa. Optical transmission for the electrode reliably decreases to a minimum at the end of the discharge and then recovers to a maximum value at the end of the charge process. In Fig. S8, operando optical spectra in 0.5 V intervals are shown for the discharge processes of (b) cycle 26 and (e) cycle 55 and for the charge processes of (c) cycle 26 and (f) cycle 55. Mirroring the results in the main text, the cyclic transmittance is also seen for every cycle monitored here.

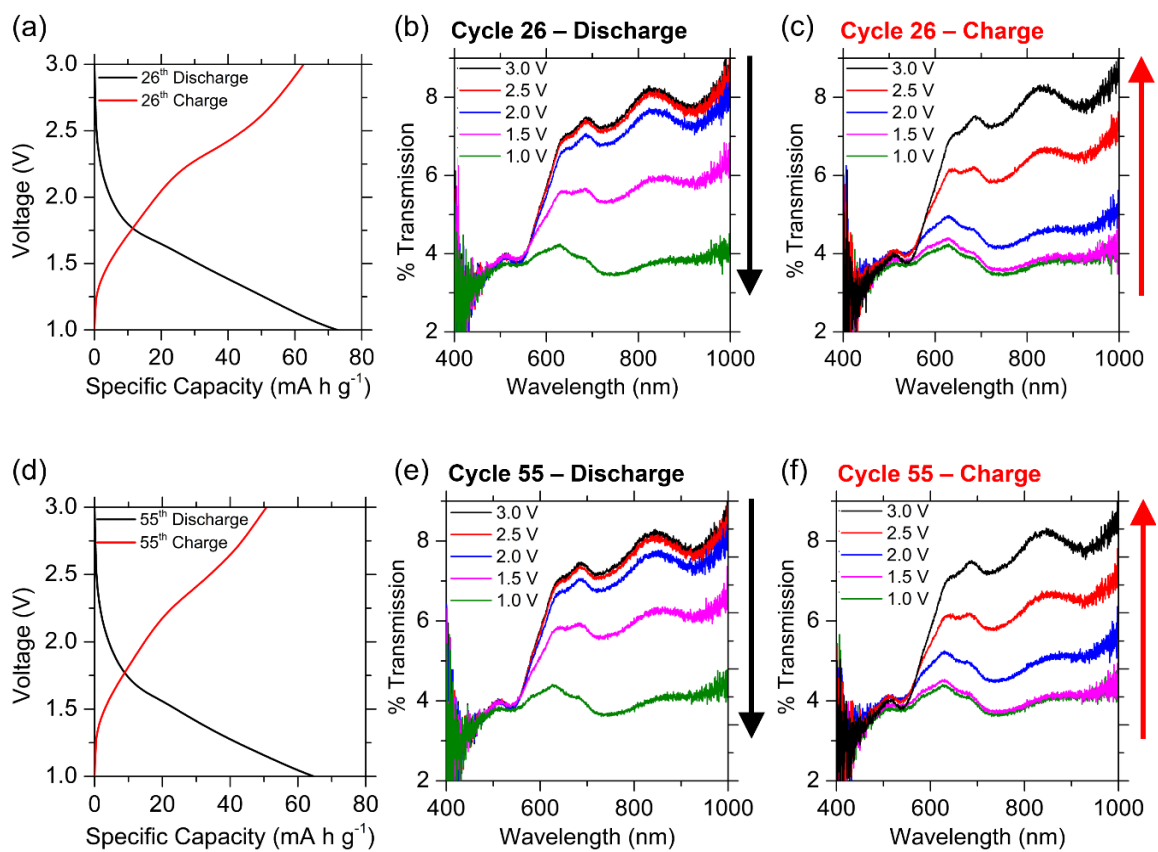

**Fig. S8** Electrochemical and optical data TiO$_2$ IOs with larger periodicities prepared using 500 nm polystyrene opal templates. Galvanostatic charge/discharge data showing calculated specific capacities for (a) cycle 26 and (d) cycle 55. Operando optical spectra recorded in 0.5 V intervals for TiO$_2$ IOs undergoing discharge from 3.0 – 1.0 V showing a decrease in overall transmittance with increasing lithiation (matching the effects depicted in the main text) for (b) cycle 26 and (e) cycle 55. The accompanying operando optical spectra for the charge process of the TiO$_2$ IO from 1.0 – 3.0 V showing a recovery in transmittance as delithiation occurs for (c) cycle 26 and (f) cycle 55.

## 5. Additional Material Characterisation for TiO$_2$ IO Electrodes

The requirement to use a custom electrode is explained above in the text accompanying Fig. S1. In short, conductive oxide substrates blacken, lowering the overall electrode transmittance significantly, with the onset of lithiation. For the results presented in this work, a custom electrode was developed as a means of avoiding this effect. A non-conductive glass substrate was used for IO formation and 20 nm of Ni metal was used to provide a conductive layer for the electrode. The poor conductivity of the substrate also affected the ability to image the IO material via SEM. Microscopy images were prone to excessive charging effects and poor image definition as a direct consequence of the poor conductivity. As a result of this effect, many of the images shown of the TiO$_2$ IO electrodes throughout the text already feature the 20 nm of Ni coating used to bestow a conductive layer to the electrode. The IOs depicted in these images feature noticeably thicker IO walls when compared to the uncoated TiO$_2$ IOs in pristine condition.

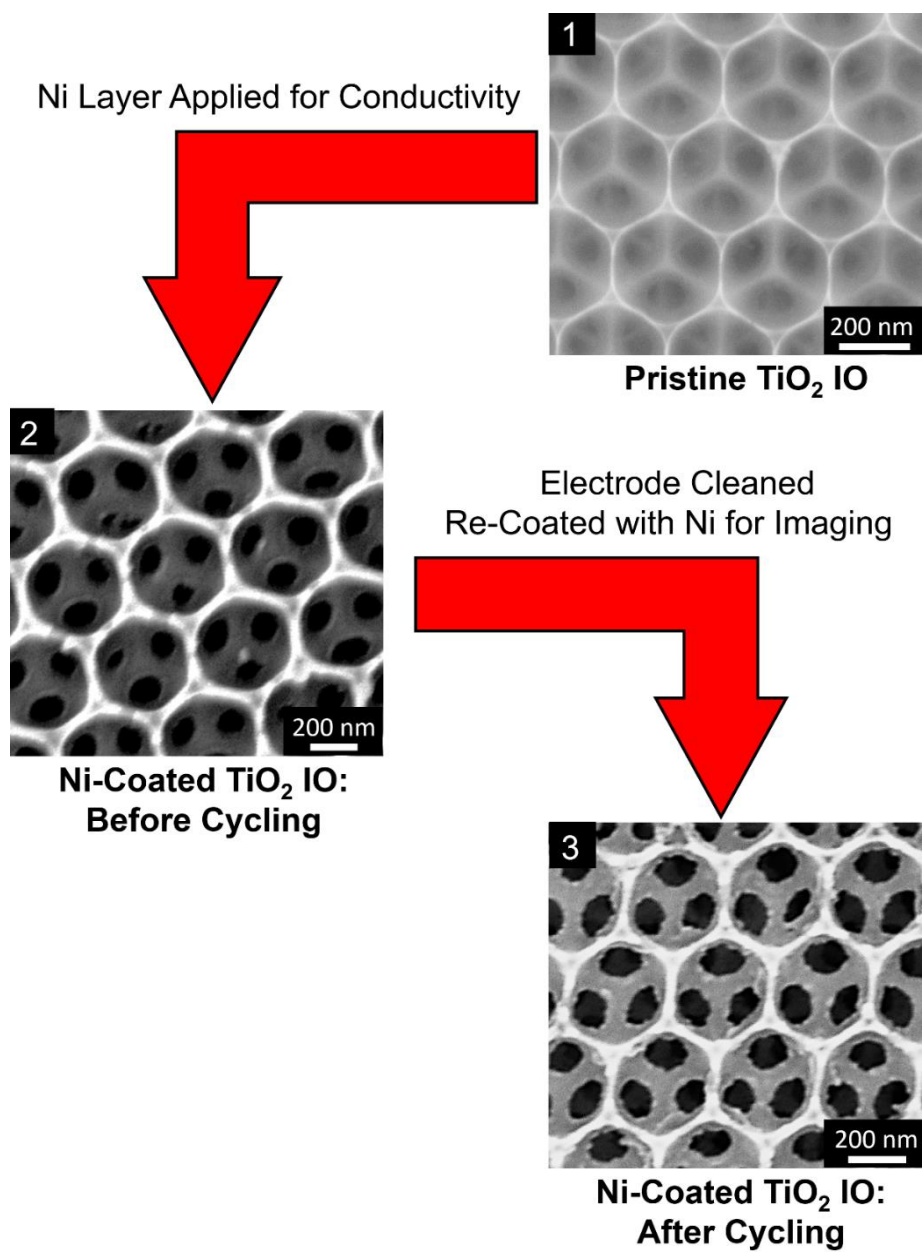

**Fig. S9** Outline of the procedure used for depositing metal coats and imaging the TiO$_2$ IO electrodes. To provide a conductive layer, 20 nm of Ni metal is first deposited onto a pristine TiO$_2$ IO. After electrochemical cycling the electrode was cleaned to remove residual electrolyte and re-coated with 20 nm of Ni metal to provide a conductive layer for SEM imaging.

Figure S9 illustrates the metal coating procedure used to image samples in this study. First the pristine TiO$_2$ IO is coated with 20 nm of Ni to provide conductivity for microscopy imaging, simultaneously fulfilling the requirement for a conductive current collector layer for the electrode. The pristine TiO$_2$ IO shown here, is a typical TiO$_2$ IO prepared on an FTO substrate using the same method and materials as the glass substrate; the conductive layer of the FTO substrate makes SEM imaging possible here. Finally, in order to image the electrode post-electrochemical cycling, the electrode needed to be cleaned (to remove residual electrolyte salt, see below for more details) and re-coated with 20 nm of Ni metal for imaging. Post-electrochemical cycling, the TiO$_2$ IOs retain their structural order yet appear visually different with rougher IOs walls and small flecks of material present throughout. It is unclear whether or not this a direct result of electrochemical cycling or as a result of the cleaning process and repeated metal layers deposited across the surface. Regardless, the periodic structure of the IOs is retained which is essential for observing the photonic stopband.

There are several different crystal phases of TiO$_2$ which can form when crystallising the material; all of these different phases of TiO$_2$ have distinctly different electrochemical and optical properties. Consequently, it is important to identify which phase of TiO$_2$ is produced when conducting analysis of the material. Voltage plateaus in galvanostatic cycling and reaction peaks in CV cycling are dependent on the type of TiO$_2$ being used. The reaction peaks observed in the CV cycles in the main text suggested the presence of anatase phase TiO$_2$[7] (and possibly bronze phase TiO$_2$[8]) when compared to literature sources working with these materials. The reported refractive indices of different TiO$_2$ phases are significantly different, thus affecting the expected position of the photonic stopband of any potential TiO$_2$ IOs. For instance, rutile phase TiO$_2$ IO is reported as having a refractive index of 2.62 at 633 nm[9] while anatase phase TiO$_2$ is reported as having a refractive index of 2.48 at 633 nm[9]. Our previous works with TiO$_2$ IOs have suggested the dominant presence of anatase phase TiO$_2$ IO using a combination of Raman spectroscopy and the expected position of the photonic stopband[10][11][2].

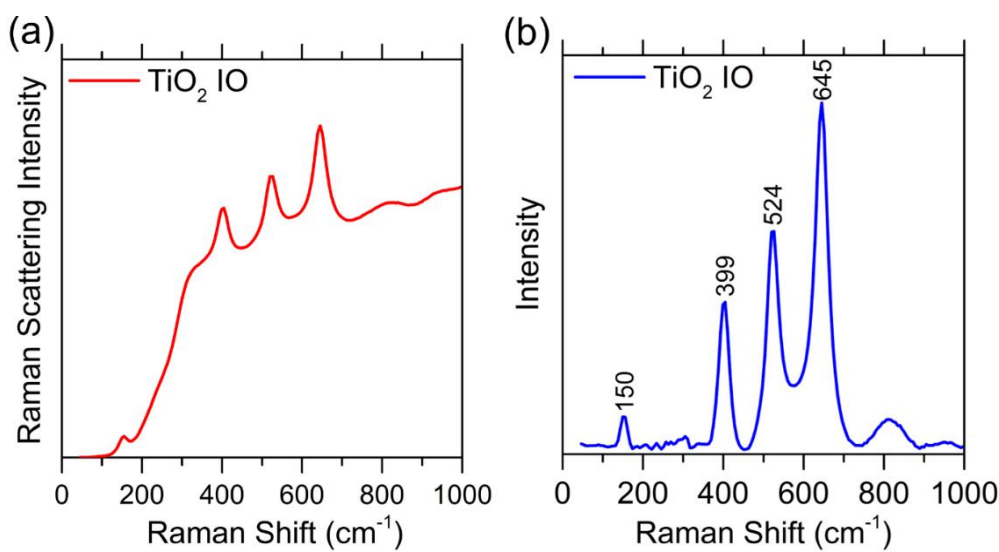

**Fig. S10** Raman scattering spectra for typical TiO$_2$ IOs prepared in this study. (a) The raw, unprocessed spectrum shows evidence of background fluorescence but clear peaks are still observed. (b) Background subtraction for the Raman spectra allows clear peaks to be observed with labelled Raman shift positions.

Figure S10 shows the Raman spectra recorded from a typical TiO$_2$ IO electrode used in this study. Raman spectra are recorded from uncoated TiO$_2$ IOs prepared on glass substrates, prior to any deposition of metal or modification of the structure. The Raman spectrum shown in Fig. S10 (a) features several prominent peaks yet is prone to background fluorescence, a common issue with Raman spectra recorded with visible laser excitation sources. Figure S10 (b) shows the same spectrum following a background subtraction to show the peaks more clearly. The Raman shift peaks for the TiO$_2$ IO material observed in this case are labelled and match with literature peaks for anatase phase TiO$_2$ and *not* rutile phase TiO$_2$[12][13].

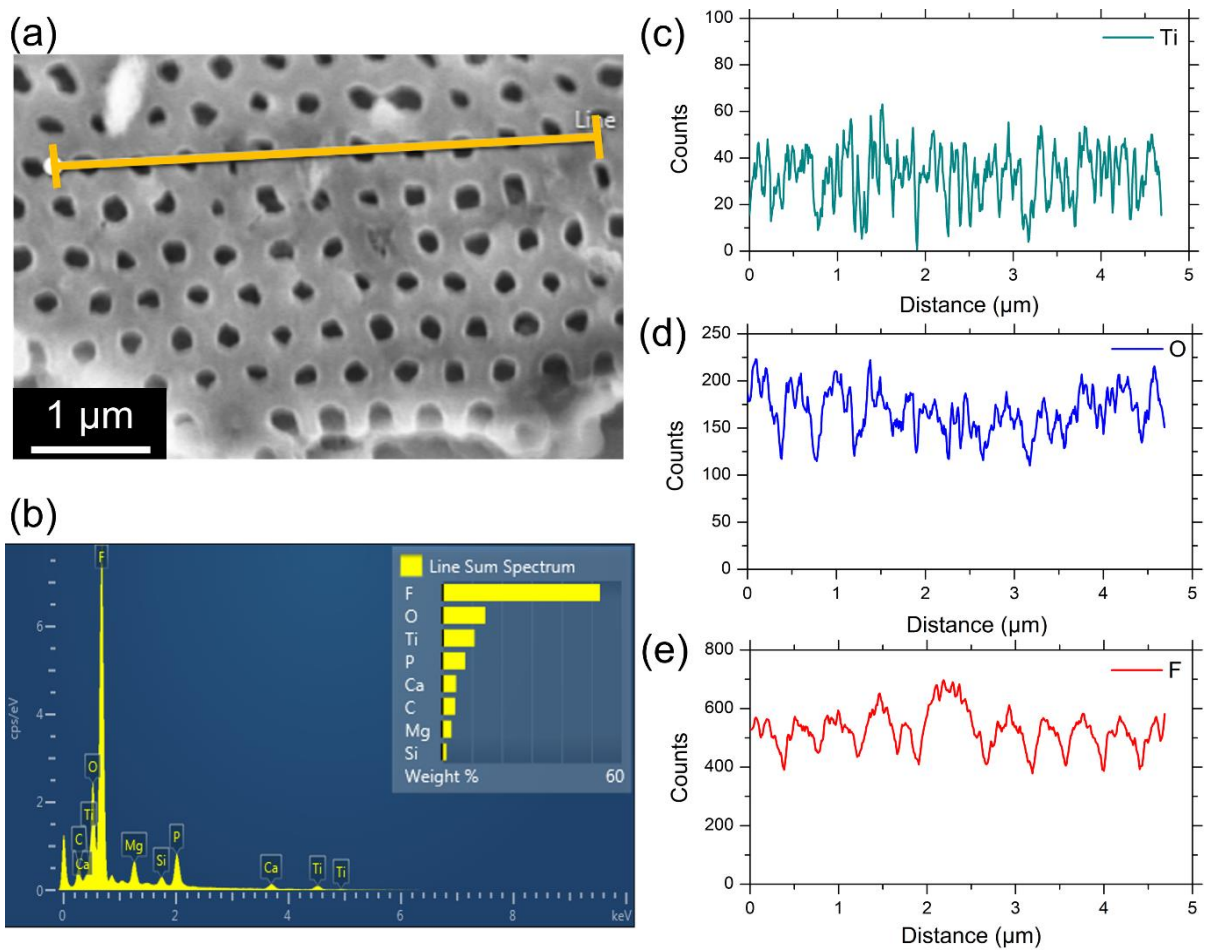

**Fig. S11** (a) An SEM image showing a line-scan across a region heavily contaminated with electrolyte salt post-electrochemical cycling with an insufficient electrode cleaning procedure. (b) The corresponding EDX elemental analysis for the line-scan region showing an abundance of elemental fluorine. Individual line-scan data tracking elemental (c) titanium, (d) oxygen and (e) fluorine shows a prominent presence of fluorine across the sample surface.

Imaging the $TiO_2$ IO electrode post-electrochemical cycling can be complicated due to the presence of residual electrolyte. An example of this effect can be seen in Fig. S11 (a) where an excess of electrolyte salt is visible on the IO surface of the SEM image of a $TiO_2$ IO electrode which was only minimally treated post-cycling. Elemental analysis performed using an EDX line-scan across a region of this electrolyte-contaminated surface reveals high levels of fluorine and phosphorous in the elemental composition (from $LiPF_6$), seen in Fig. S11 (b). Several line-scans for titanium, oxygen and fluorine are shown in Figs. S11 (c), (d) and (e), respectively. For a $TiO_2$ IO electrode, high content of elemental Ti and O should be expected. The line-scan counts show a dominant presence of elemental F, suggesting an excess of residual electrolyte.

To prevent the obscuring of the IO surface with residual electrolyte when imaging the surface, it is essential to establish that the cleaning of the electrode removes the majority of material not related to the IO material. This is an important consideration to ensure that SEM images captured of the electrode post-cycling are not complicated with other material and just show the appearance of the IO material. For the $TiO_2$ IO electrodes used in this work, allowing the electrode to sit in a vial of de-ionised water for 3 – 4 hours was found to be sufficient to remove the majority of residual electrolyte from the surface of the electrode.

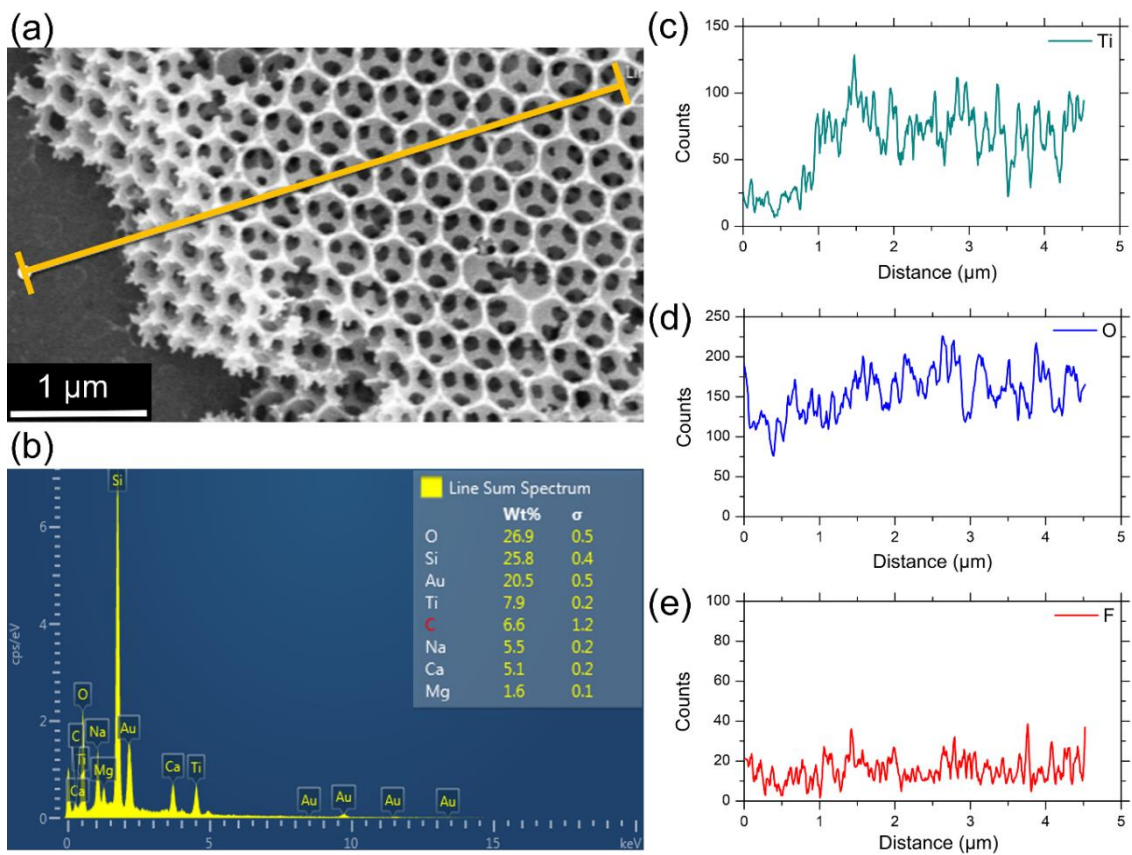

**Fig. S12** (a) An SEM image showing a line-scan across a region which has been carefully cleaned to remove residual electrolyte post-electrochemical cycling. (b) The accompanying EDX elemental analysis does not detect a significant presence of fluorine in this case. Individual line-scan data tracking elemental (c) titanium, (d) oxygen and (e) fluorine detects a minimal concentration of fluorine when the sample has undergone sufficient post-treatment to remove residual electrolyte.

Figure S12 (a) shows the appearance of a typical $TiO_2$ IO electrode post-cycling which has been cleaned using this process. The structure of the IO material is maintained and resembles its initial appearance prior to cycling. Importantly, there is no visible presence of residual electrolyte contaminating the surface of this SEM image. Elemental analysis of a line-scan region across this SEM images does not appear to detect any significant levels of fluorine or phosphorous for this area, suggesting a lack of $LiPF_6$ electrolyte salt, as seen in the EDX elemental composition in Fig. S12 (b). EDX line-scans of elemental titanium, oxygen and fluorine content are shown for this region in Figs. 12 (c), (d) and (e), respectively. Compared to the electrolyte contaminated region in Fig. S11, the region depicted here has minimal presence of fluorine and shows high counts of titanium and oxygen, as expected from a $TiO_2$ IO. These results would suggest a very minor presence of residual electrolyte remaining on the cleaned electrode sample. Consequently, any SEM images of a $TiO_2$ IO electrode post-cycling is this study adopted this electrode cleaning process to remove residual electrolyte.

## 6. References

1. Ou, J., Wu, S., Yang, L. & Wang, H. Facile Preparation of NiO@graphene Nanocomposite with Superior Performances as Anode for Li-ion Batteries. *Acta Metallurgica Sinica (English Letters)* **35**, 212-222 (2022). https://doi.org:10.1007/s40195-021-01283-5
2. Lonergan, A., Murphy, B. & O'Dwyer, C. Photonic Stopband Tuning in Metallo-Dielectric Photonic Crystals. *ECS Journal of Solid State Science and Technology* **10**, 085001 (2021). https://doi.org:10.1149/2162-8777/ac19c5
3. McNulty, D., Carroll, E. & O'Dwyer, C. Rutile TiO2 Inverse Opal Anodes for Li-Ion Batteries with Long Cycle Life, High-Rate Capability, and High Structural Stability. *Advanced Energy Materials* **7**, 1602291 (2017). https://doi.org:https://doi.org/10.1002/aenm.201602291
4. Wu, N. *et al.* Anatase inverse opal TiO2-x@N-doped C induced the dominant pseudocapacitive effect for durable and fast lithium/sodium storage. *Electrochimica Acta* **299**, 540-548 (2019). https://doi.org:https://doi.org/10.1016/j.electacta.2019.01.040
5. Lu, R., Yang, A., Xue, Y., Xu, L. & Zhu, C. Analysis of the key factors affecting the energy efficiency of batteries in electric vehicle. *World Electric Vehicle Journal* **4**, 9-13 (2010). https://doi.org:10.3390/wevj4010009
6. Wang, S. *et al.* in *Battery System Modeling* (eds Shunli Wang *et al.*) 1-46 (Elsevier, 2021).
7. Gao, X. *et al.* Preparation and Electrochemical Characterization of Anatase Nanorods for Lithium-Inserting Electrode Material. *The Journal of Physical Chemistry B* **108**, 2868-2872 (2004). https://doi.org:10.1021/jp036821i
8. Laskova, B., Zukalova, M., Zukal, A., Bousa, M. & Kavan, L. Capacitive contribution to Li-storage in TiO2 (B) and TiO2 (anatase). *Journal of Power Sources* **246**, 103-109 (2014). https://doi.org:https://doi.org/10.1016/j.jpowsour.2013.07.073
9. Möls, K. *et al.* Influence of phase composition on optical properties of TiO2: Dependence of refractive index and band gap on formation of TiO2-II phase in thin films. *Optical Materials* **96**, 109335 (2019). https://doi.org:https://doi.org/10.1016/j.optmat.2019.109335
10. Lonergan, A., McNulty, D. & O'Dwyer, C. Tetrahedral framework of inverse opal photonic crystals defines the optical response and photonic band gap. *Journal of Applied Physics* **124**, 095106 (2018). https://doi.org:10.1063/1.5033367
11. Lonergan, A., Hu, C. & O'Dwyer, C. Filling in the gaps: The nature of light transmission through solvent-filled inverse opal photonic crystals. *Physical Review Materials* **4**, 065201 (2020). https://doi.org:10.1103/PhysRevMaterials.4.065201
12. Zhang, W. F., He, Y. L., Zhang, M. S., Yin, Z. & Chen, Q. Raman scattering study on anatase TiO2 nanocrystals. *Journal of Physics D: Applied Physics* **33**, 912 (2000). https://doi.org:10.1088/0022-3727/33/8/305
13. Zhang, J., Li, M., Feng, Z., Chen, J. & Li, C. UV Raman Spectroscopic Study on TiO2. I. Phase Transformation at the Surface and in the Bulk. *The Journal of Physical Chemistry B* **110**, 927-935 (2006). https://doi.org:10.1021/jp0552473